\documentclass{article}

\usepackage[preprint]{neurips_2025}
\usepackage{alltt}
\usepackage{tabularx}
\usepackage{graphicx}
\usepackage{booktabs}
\usepackage{xcolor}
\usepackage{colortbl}
\usepackage{array}
\usepackage{float}
\usepackage{enumitem}
\usepackage{makecell}
\usepackage{tikz}
\usepackage{pgfplots}
\usepackage{hyperref}
\usepackage[toc,page]{appendix}
\usepackage{subcaption}
\usepgfplotslibrary{groupplots}
\usepackage{pgf}
\usepackage{makecell}
\usepackage{booktabs}
\usepackage{xcolor}
\usepackage{comment}
\usepackage{colortbl}
\usepackage{tcolorbox}
\usepackage{geometry}
\usepackage{pifont}
\definecolor{darkgreen}{RGB}{34,139,34}
\newcommand{\cmark}{\textcolor{darkgreen}{\ding{51}}}  
\newcommand{\xmark}{\textcolor{purple}{\ding{55}}}        

\definecolor{lightred}{rgb}{1,0.85,0.85}

\newcommand{\shirwu}[1]{\textcolor{teal}{(Shirley: #1)}}
\usepackage[utf8]{inputenc} 
\usepackage[T1]{fontenc}    
\usepackage{hyperref}       
\usepackage{url}            
\usepackage{booktabs}       
\usepackage{amsfonts}       
\usepackage{nicefrac}       
\usepackage{microtype}      
\usepackage{xcolor}         
\newcommand{\xhdr}[1]{{\noindent\bfseries #1}.}
\newcommand{\ours}{Finance Agent Benchmark}
\title{\ours{}: Benchmarking LLMs on Real-world Financial Research Tasks}

\author{%
  Antoine Bigeard \\
  Vals AI, Inc. \\
  \texttt{antoine@vals.ai} \\
  \And
  Rayan Krishnan \\
  Vals AI, Inc. \\
  \texttt{rayan@vals.ai} \\
    \And
  Shirley Wu \\
  Stanford University \\
  \texttt{shirwu@stanford.edu} \\
  \And
  Langston Nashold \\
  Vals AI, Inc. \\  
  \texttt{langston@vals.ai}
}

\begin{document}

\maketitle

\vspace{-20pt}
\begin{abstract}


Artificial Intelligence (AI) technology has emerged as a transformative force in financial analysis and the finance industry, though significant questions remain about the full capabilities of Large Language Model (LLM) agents in this domain.
We present the Finance Agent Benchmark, featuring challenging and diverse real-world finance research problems which require LLMs to perform complex analysis with the use of of recent SEC filings.
We construct the benchmark using a taxonomy of nine financial task categories, developed in consultation with experts from banks, hedge funds, and private equity firms. The dataset includes 537 expert-authored questions,  covering tasks from information retrieval to complex financial modeling, where each question was validated through a rigorous review process to ensure accuracy and relevance.
Moreover, we implement an agentic harness that equips LLMs with tools sufficient to produce an accurate response, including as Google Search and EDGAR database access.
Overall, Finance Agent Benchmark provides a comprehensive testbed for measuring the progress of LLM-driven finance agents. Our evaluation reveals significant limitations in current AI capabilities—even the best-performing model (OpenAI's o3) achieved only 46.8\% accuracy, at an average cost of \$3.79 per query. It underscores the need for further advancements before reliable deployment in high-stakes finance settings.

\end{abstract}

\vspace{-10pt}
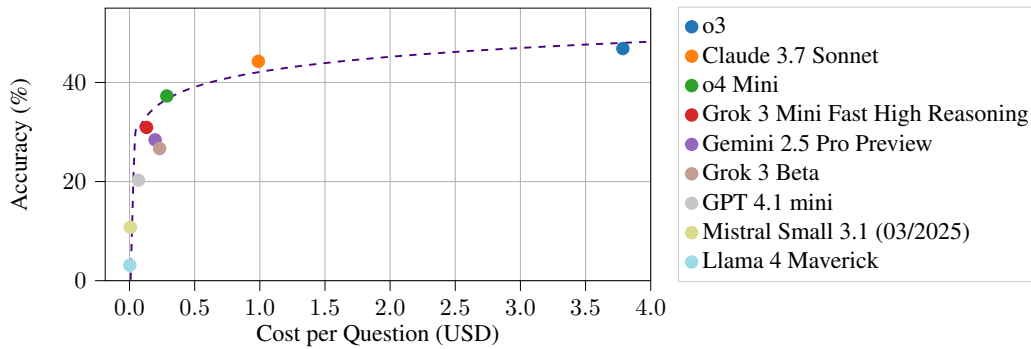
\begin{figure}[h]
    \begin{centering}
    \resizebox{\textwidth}{!}{%
\begin{tikzpicture}
\definecolor{crimson2143940}{RGB}{214,39,40}
\definecolor{darkgray176}{RGB}{176,176,176}
\definecolor{darkorange25512714}{RGB}{255,127,14}
\definecolor{forestgreen4416044}{RGB}{44,160,44}
\definecolor{khaki219219141}{RGB}{219,219,141}
\definecolor{lightblue158218229}{RGB}{158,218,229}
\definecolor{lightgray204}{RGB}{204,204,204}
\definecolor{mediumpurple148103189}{RGB}{148,103,189}
\definecolor{pink247182210}{RGB}{247,182,210}
\definecolor{rosybrown196156148}{RGB}{196,156,148}
\definecolor{silver199}{RGB}{199,199,199}
\definecolor{steelblue31119180}{RGB}{31,119,180}
\definecolor{logcurvecolor}{RGB}{75,0,130} 

\begin{axis}[
legend cell align={left},
legend style={
  fill opacity=1,
  draw opacity=1,
  text opacity=1,
  at={(1.05,0.5)},
  anchor=west,
  draw=lightgray204
},
width=8cm,
height=4cm,
scale only axis=true,
tick align=outside,
tick pos=left,
x grid style={darkgray176},
xlabel={Cost per Question (USD)},
xmajorgrids,
xmin=-0.186848741076939, xmax=4,
xtick style={color=black},
xtick={0,0.5,1,1.5,2,2.5,3,3.5,4},
xticklabels={
  \(\displaystyle 0.0\),
  \(\displaystyle 0.5\),
  \(\displaystyle 1.0\),
  \(\displaystyle 1.5\),
  \(\displaystyle 2.0\),
  \(\displaystyle 2.5\),
  \(\displaystyle 3.0\),
  \(\displaystyle 3.5\),
  \(\displaystyle 4.0\)
},
y grid style={darkgray176},
ylabel={Accuracy (\%)},
ymajorgrids,
ymin=0, ymax=55,
ytick style={color=black}
]
\addplot [semithick, steelblue31119180, opacity=1, mark=*,mark size=2.5, mark options={solid}, only marks]
table {%
3.78610164342291 46.837344874371
};
\addlegendentry{o3}
\addplot [semithick, darkorange25512714, opacity=1, mark=*,mark size=2.5, mark options={solid}, only marks]
table {%
0.988625833099283 44.2639091676131
};
\addlegendentry{Claude 3.7 Sonnet}
\addplot [semithick, forestgreen4416044, opacity=1, mark=*,mark size=2.5, mark options={solid}, only marks]
table {%
0.2862863155701 37.276820487755
};
\addlegendentry{o4 Mini}
\addplot [semithick, crimson2143940, opacity=1, mark=*,mark size=2.5, mark options={solid}, only marks]
table {%
0.13049179427344 30.9238811654442
};
\addlegendentry{Grok 3 Mini Fast High Reasoning}
\addplot [semithick, mediumpurple148103189, opacity=1, mark=*,mark size=2.5, mark options={solid}, only marks]
table {%
0.196306608510841 28.433138978131
};
\addlegendentry{Gemini 2.5 Pro Preview}
\addplot [semithick, rosybrown196156148, opacity=1, mark=*,mark size=2.5, mark options={solid}, only marks]
table {%
0.230873824907237 26.6630122657041
};
\addlegendentry{Grok 3 Beta}
\addplot [semithick, silver199, opacity=1, mark=*,mark size=2.5, mark options={solid}, only marks]
table {%
0.0683812668820963 20.2872449332544
};
\addlegendentry{GPT 4.1 mini}
\addplot [semithick, khaki219219141, opacity=1, mark=*,mark size=2.5, mark options={solid}, only marks]
table {%
0.00605787514245645 10.773305418569
};
\addlegendentry{Mistral Small 3.1 (03/2025)}
\addplot [semithick, lightblue158218229, opacity=1, mark=*,mark size=2.5, mark options={solid}, only marks]
table {%
0.00233937247067305 3.11206235775217
};
\addlegendentry{Llama 4 Maverick}

\addplot [
    thick,
    color=logcurvecolor,
    dashed,
    domain=0.00001:4,
    samples=100,
    smooth
]
{4.4*ln(x)+42.13};

\end{axis}
\end{tikzpicture}
    }
    \vspace{-15pt}
    \caption{Cost-Accuracy pareto curve results on \ours{}. The Finance Agent Benchmark reveals a clear logarithmic relationship between accuracy and cost, with a sharp diminishing return beyond \$1 USD per question—highlighting that even today's most sophisticated models struggle to achieve greater than 50\% accuracy on real-world financial tasks.}
    \label{fig:pareto_curve_subset}
    \end{centering}
\end{figure}

\vspace{-10pt}

\section{Introduction}



The finance sector is a critical domain with large economic impact and operational scale. With daily foreign exchange turnover averaging \$7.5 trillion \cite{bis2022fxturnover}, financial operations demand extensive human resources for routine tasks. Entry-level finance professionals sometimes spend up to 40\% of their workweek on gathering data rather than analyzing it \cite{pwc_effectiveness_benchmark}—time that could be better used for strategic analysis that drives business value. The repetitive and essential nature of these tasks requires significant workforce investment and creates inefficiencies.

Recent developments in autonomous AI agents offer a promising solution for financial workflows \cite{yehudai2025surveyevaluationllmbasedagents, kapoor2024aiagents}. These systems have shown advancements in their ability to perform complex tasks that involve unstructured data and multi-step reasoning—capabilities essential for financial analysis \cite{an2024finverseautonomousagentversatile}. By automating routine yet time-consuming tasks, AI agents can greatly reduce human effort. However, despite this clear potential, there remains a need for robust, domain-specific benchmarks to assess the capabilities of financial agents.

Previous benchmarks often fail to address the interactive environments and nuanced reasoning an agent must undergo when completing industry-specific tasks, leaving performance in real-world scenarios uncertain \cite{chen2025standardbenchmarks}. This deficiency is particularly concerning given that AI models can generate plausible-sounding misinformation and hallucinations \cite{williamson2024hallucinations}, which, if relied upon for high-stake, high-impact decisions, could lead to severe negative outcomes.

For instance, if an AI model misinterprets a company's earnings report, incorrectly identifying a consistent earnings beat over four consecutive quarters, it could lead to misguided decision to invest in the company.
Without proper evaluation, these systems cannot be trusted for high-stakes financial applications. A dedicated benchmark is essential for tracking the state-of-the-art financial agents and highlighting the remaining deficits in model capabilities in this domain.


\begin{figure}[t]
    \centering
    \includegraphics[width=1\linewidth]{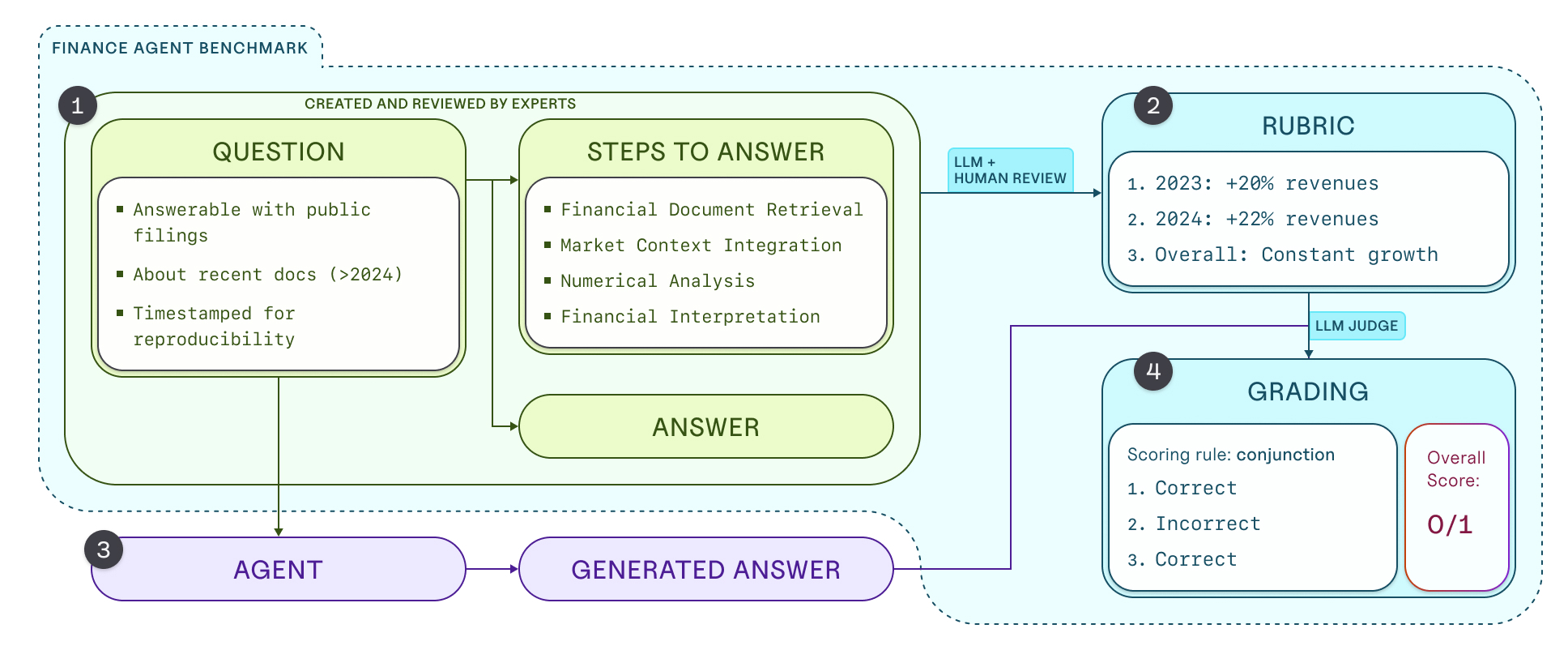}
    \caption{
    Architecture of the Finance Agent Benchmark. The framework features a structured evaluation process with four key steps: (1) Data Creation: experts identify practical and common financial questions requiring access to public financial documents and provide reference answers. (2) Rubric Development: expert-generated data is used to create robust rubrics with expected calculations and reasoning steps for standardized LLM evaluation. (3) Agent Evaluation: questions are processed through LLMs equipped with necessary tools to generate answers. (4) Answer Grading: an LLM-as-judge scoring system using LLM-as-judge applies conjunction rules to determine correctness across multiple criteria. 
    }
    \label{fig:finance-agent-benchmark}
    \vspace{-10pt}
\end{figure}






To fill this gap, we present Finance Agent Benchmark (Figure \ref{fig:finance-agent-benchmark}), a novel, standardized framework to rigorously evaluate AI agents on real-world financial analysis tasks.
Our benchmark offers:

\begin{itemize}[leftmargin=*]
    \item \textbf{Realistic queries representing real-world challenges:} We collaborated with seven domain experts across the financial services industry (banks, hedge funds, private equity) to identify and categorize common analytical tasks performed with public financial documents, see Table \ref{tab:task-types}.

    \item \textbf{Step-by-step expert annotation for reliable validation}: We constructed this comprehensive dataset with questions, corresponding answers and reasoning trajectories (steps to answer) created by the experts. Each question is verifiable through public U.S. Securities and Exchange Commission (SEC)'s  filings from the Electronic Data Gathering, Analysis, and Retrieval (EDGAR) database, official repository for public company filings, and underwent peer-review by the experts.

    \item \textbf{Multi-dimensional Evaluation Framework}: Our benchmark uses an LLM-as-judge approach with rubric-based assessment that evaluates answers against specific components of expert solutions rather than holistically. The framework incorporates multiple accuracy verification checks and a contradiction detection mechanism to identify factual inconsistencies between generated and expert answers.
    
    \item \textbf{Model-Agnostic Evaluation Harness for Benchmark Baseline}: We developed an agentic evaluation framework (Figure \ref{fig:agentic-harness-graph}) equipped with multiple tools, including Google Search and EDGAR Research Filing Search, enabling standard benchmarking and performance assessment across different LLMs. We established performance baselines using our evaluation harness, providing a foundation for future research in finance-focused AI agents.
\end{itemize}

Our analysis  in Figure \ref{fig:pareto_curve_subset} indicates that substantial progress is still needed—the best-performing baseline model, o3, achieved only 46.8\% accuracy. This highlights the benchmark’s value in testing the ability of models to perform time-intensive tasks expected of entry level finance analysts. Notably, no model surpassed 50\% accuracy, underscoring the advancements required before these systems can be deployed autonomously and reliably in the financial sector. 

Nevertheless, these models demonstrate notable efficiency advantages: even the most expensive model (o3) averages just 3.1 minutes per task and costs \$3.78, compared to human experts requiring 16.8 minutes and costing \$25.66 \footnote{These calculations only account for time spent directly answering the question, and exclude the time to create the question, review the answer, etc.} for equivalent analyses. This efficiency-to-performance ratio points to the promising potential of these models in supporting human analysts, even as work continues toward improving their accuracy for more autonomous applications.



\section{\ours{}: High-quality Financial Benchmark}


\subsection{Dataset Creation Process}

\textbf{Expert Consultation and Taxonomy Development.}
We collaborated with financial industry experts from banks, hedge funds, and startups to identify common analytical tasks and develop a comprehensive taxonomy of finance analysis questions. Figure \ref{fig:finance-agent-benchmark} shows the overall development process of the benchmark. The experts had at least 2-3 years of working experience at bulge bracket firms such as Goldman Sachs and J.P. Morgan. The resulting taxonomy categorizes tasks based on complexity and frequency in real investment banking contexts, ranging from basic retrieval to complex market analysis. See dataset taxonomy in Table \ref{tab:task-types}.

\textbf{Question Generation and Quality Control.}
The experts then crafted 537 questions across nine task categories, with instructions to create queries requiring multiple financial documents. Each entry includes the question, ground-truth answer, necessary source documents, and a step-by-step solution approach. Questions were designed to be answerable without additional information beyond what is available on the open internet or in public filings. The experts focused their question generation on documents published no earlier than 2024 (after most training cutoffs).

\textbf{Rigorous Validation Process.}
Each question underwent peer review by a different expert to verify calculations and content validity. Questions containing errors were either corrected or removed. The authors of this paper conducted a final review to standardize formatting and ensure consistency throughout the benchmark. This multi-stage validation process ensured that our dataset  represents real-world financial analysis challenges.

\begin{table}[t]
    \centering
    \small
    \renewcommand{\arraystretch}{1.1}
    \begin{tabular}{p{0.20\textwidth}p{0.5\textwidth}p{0.1\textwidth}p{0.10\textwidth}}
    \toprule
    \textbf{Task Name} & \textbf{Question Example} & \textbf{Difficulty} & \textbf{Count} \\
    \midrule
    Quantitative Retrieval & \textit{What was the quarterly revenue of Salesforce (NYSE:CRM) for the quarter ended December 31, 2024?} & Easy & 102 (19\%) \\
    \rowcolor[gray]{0.95}
    Qualitative Retrieval & \textit{Describe the product offerings and business model of Microsoft (NASDAQ:MSFT)?} & Easy & 97 (18\%) \\
    Numerical Reasoning & \textit{What is \% of revenue derived from AWS in each year and the 3 year CAGR from 2021-2024 of Amazon?} & Easy & 83 (15\%) \\
    \rowcolor[gray]{0.95}
    Complex Retrieval & \textit{Please briefly summarize the most recent capital raise conducted by Viking Therapeutics (NASDAQ:VKTX).} & Medium & 29 (6\%) \\
    Adjustments & \textit{What is Lemonade Insurance’s Adjusted EBITDA for the year ended December 31, 2024?} & Medium & 43 (8\%) \\
    \rowcolor[gray]{0.95}
    Beat or Miss & \textit{How did Lam Research’s revenue compare to management projections (at midpoint) on a quarterly basis in 2024? Format as \% BEAT or MISS. Use guidance provided on a quarterly basis.} & Medium & 69 (13\%) \\
    Trends & \textit{Which Geographic Region has Airbnb (NASDAQ: ABNB) experienced the most revenue growth from 2022 to 2024?} & Hard & 33 (6\%) \\
    \rowcolor[gray]{0.95}
    Financial Modeling & \textit{How much M\&A firepower does Amazon have as of FY2024 end including balance sheet cash, non-restricted cash and other short term investments, and up to 2x GAAP EBITDA leverage? Round to nearest billion.} & Hard & 47 (9\%) \\
    Market Analysis & \textit{Compare the quarterly revenue growth of FAANG companies between 2022-2024.} & Hard & 34 (6\%) \\
    \bottomrule
    \end{tabular}
    \caption{Task Types and Analyst Difficulty Levels
    }
    \label{tab:task-types}
\end{table}

\subsection{Evaluation Methodology}

The LLM-as-judge methodology is becoming more widespread, as the complexity of evaluation scales with the complexity of tasks studied \cite{liu2023GEvalNE}. Our dataset employs a refined LLM-as-judge approach designed to maximize accuracy by systematically reducing both false positives and false negatives. We particularly focus on false negatives, which tend to occur more frequently when using language models as evaluators \cite{pathak2025rubricneedenhancingllmbased}. 

\paragraph{Rubric-Based Evaluation.}
Rather than evaluating answers holistically, we separate assessment into distinct rubrics that align with specific key points from expert answers \cite{bai2022constitutionalaiharmlessnessai}. For example, if an expert's answer contains three main points, we check for the presence of each point in the generated answer separately. For straightforward responses (e.g., numerical values like ``\$5,234,183``), direct comparison methods are sufficient. This structured approach enhances evaluation reliability by ensuring thorough coverage of all critical aspects while maintaining high standards of factual accuracy and completeness.

\paragraph{Rubric Generation Process.}
We used GPT-4o to automatically extract initial rubrics from the experts' questions and answers (see Appendix \ref{appendix:instruction-prompt} for the exact prompts used). Each question, answer, and its associated rubrics were then manually reviewed by the authors of this paper to verify they had accurately captured the key points of the answer written by the expert.

\paragraph{Contradiction Detection.}
To ensure factual accuracy, we implement a dedicated ``contradiction rubric'' that examines whether any part of the generated answer conflicts with the expert's answer. This approach leverages the observation that identifying contradictions between texts is typically more reliable than confirming complete agreement on all points \cite{thakur2025judgingjudgesevaluatingalignment}. An example is provided in Appendix \ref{appendix:evaluation-example}.

\paragraph{Dataset Split.}
\label{section:dataset-split}
We divided the dataset into three parts: a public validation set (50 samples under the CC BY 4.0 License, available in the project's repository), a private validation set (150 samples available to researchers upon request), and a test set (337 samples kept fully private to prevent future overfitting and contamination issues \cite{xu2024benchmarkdatacontaminationlarge}). All metrics reported in this paper were calculated on the complete 537 samples. Further details about the split are available in Appendix \ref{appendix:dataset-split}.




\section{Financial Agent Harness
}
\begin{figure}[t]  
    \centering
    \includegraphics[width=1\linewidth]{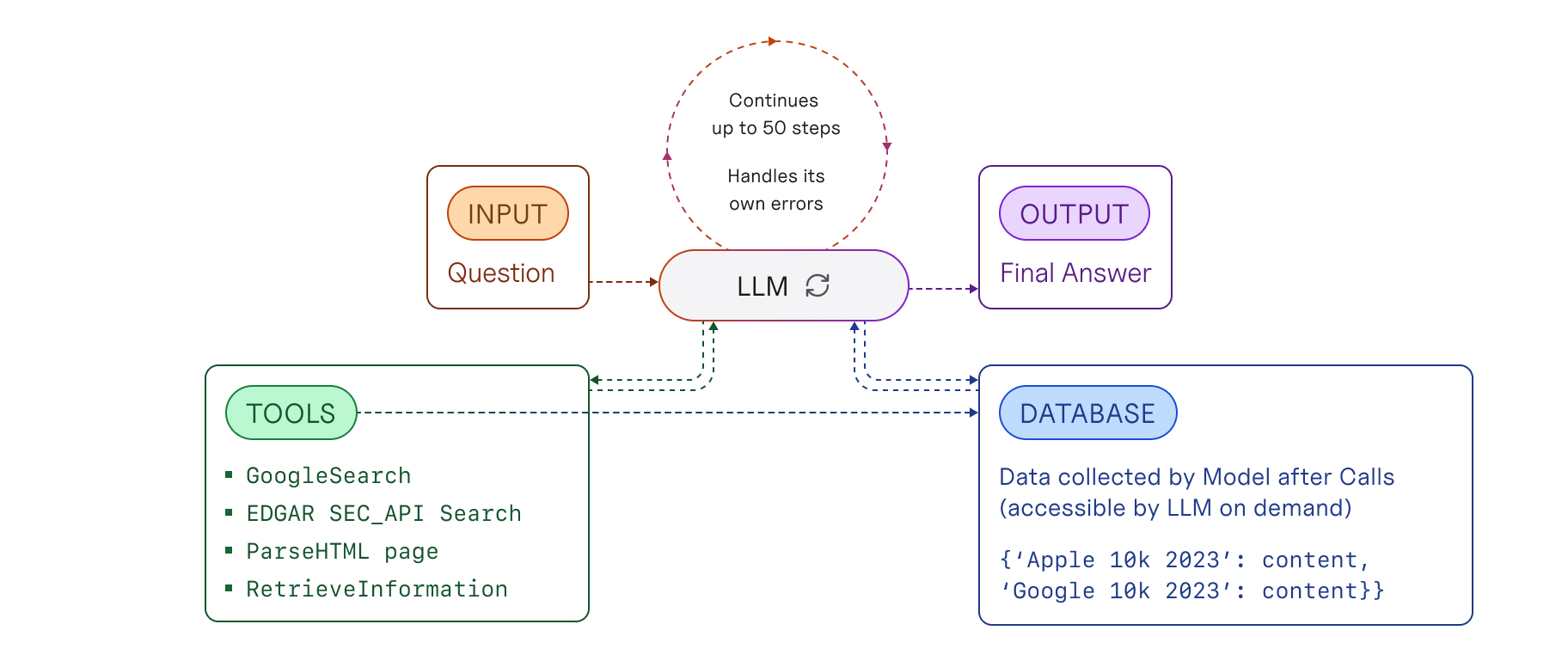}
    \vspace{-10pt}
    \caption{Financial Agent Harness architecture showing the interaction flow between LLM components, specialized tools, and information sources.}
    \label{fig:agentic-harness-graph}
\end{figure}

We created an agentic harness\footnote{\url{https://doi.org/10.5281/zenodo.15428823}} for the LLMs shown in Figure \ref{fig:agentic-harness-graph} to produce the generations being evaluated. We built this infrastructure based on commonly used frameworks like ReAct \cite{yao2023reactsynergizingreasoningacting} \cite{liu2023agentbenchevaluatingllmsagents} \cite{wang2024gtabenchmarkgeneraltool}, and informed by recent agentic evaluation benchmarks such as PaperBench \cite{starace2025paperbenchevaluatingaisability} and SWE-Lancer \cite{miserendino2025swelancerfrontierllmsearn}. This harness allows the models to be evaluated in an environment where they had access to a set of tools that is sufficient to produce an accurate response.

\subsection{Tools Overview}

The harness employs a variety of specialized tools to access and process financial information. Each tool is designed to perform a specific function:

\begin{itemize}[leftmargin=*]
\item \textbf{GoogleSearch}: A tool to access general web search.
\item \textbf{EdgarSearch}: A tool to access the EDGAR database, containing public SEC filings\footnote{https://www.sec.gov/search-filings}.
\item \textbf{ParseHTML}: A tool that extracts and saves content from HTML web pages in a key-value store database.
\item \textbf{RetrieveInformation}: A tool to retrieve stored document content from the key-value database and send the retrieved content to the model.

\end{itemize}

The ParseHTML and the RetrieveInformation tools collectively allow the model to manage its own context window. The human experts did not make use of any additional tools when writing and answering their questions. More information about each tool is available in Appendix \ref{appendix:tool-use}.

\subsection{Environment Feedback}
Environment feedback is a critical aspect of agent environments \cite{yehudai2025surveyevaluationllmbasedagents}.
As described in Figure \ref{fig:agentic-harness-graph}, after any tool call, the model receives feedback on whether the call succeeded or failed, and the result of the call (on success).

In our case, most errors occurred when calling the underlying API for each tool. We categorize into two types: retryable and agent errors.

\begin{itemize}[leftmargin=*]
    \item \textbf{Retryable}: These primarily include rate limit errors from APIs. We do not treat them as agent errors; instead, we automatically retry the request using exponential backoff.
    \item \textbf{Agent errors}: These include issues such as exceeding token limits or providing incorrect arguments to tools. We consider these as model-generated errors and return them as tool responses.
\end{itemize}

Further details on our error-handling approach can be found in Appendix \ref{appendix:error-handling}.

These tools work together to enable the agent to research financial information effectively while gracefully handling various error conditions. We have released the full evaluation harness on our \href{https://github.com/vals-ai/finance-agent}{GitHub}. 

\section{Experiments and Results}

\subsection{Experiments Setup}
\label{subsection:experiments-setup}


\xhdr{Setup and Hyperparameters} All models were evaluated under consistent conditions to ensure a fair comparison:
\begin{itemize}[leftmargin=*]
    \item \textbf{Prompting:} All models were prompted identically using the instruction provided in Appendix~\ref{appendix:instruction-prompt}. No system prompt was used, as not all providers or models support this feature.
    
    \item \textbf{Temperature:} The temperature was set to 0 for all models that supported this parameter to ensure better reproducibility of results~\cite{Ouyang_2025}. For models without configurable temperature, the default temperature (typically 1) was used.
    
    \item \textbf{Token Limit:} The maximum token limit was set sufficiently high (16,384 tokens by default) to avoid truncation before response completion. In practice, most responses did not approach this limit.
    
    \item \textbf{Compute Environment:} All experiments were lightweight in terms of memory and compute, as they primarily involved API calls. They were conducted on an AWS \texttt{t2.2xlarge} EC2 instance.
\end{itemize}

\xhdr{Evaluation Metrics} We report two types of accuracy: 
\begin{itemize}[leftmargin=*]
    \item \textbf{Class-Balanced accuracy}: An average accuracy was computed for each category of question. The per-category scores were averaged, with equal weighting given to each. 
    \item \textbf{Naive accuracy}: The percentage of questions the model got right, regardless of category.
\end{itemize}

\subsection{Results}
\label{subsection:results}

We find the class-balanced accuracy to be more representative of the model's general agentic financial capabilities. This is because some retrieval categories are overrepresented and often require significantly fewer tool calls or agentic behavior. If not otherwise specified, the figures below show results on the Class-Balanced Accuracy.

\begin{table}[]
    \centering
    \resizebox{\textwidth}{!}{%
    \begin{tabular}{lrrrrc}
    \toprule
    Model & \makecell{Acc. (Class-Balanced) ↑} & \makecell{Acc. (Naive) ↑} & \makecell{Time per Query ↓} & \makecell{Cost per Query ↓} & Reasoning \\
    \midrule
    o3 & \textbf{ \textcolor{darkgreen}{46.8 ± 2.2}} & 51.4 ± 2.2 & 186.5s (3.1m) & \textbf{ \textcolor{purple}{\$3.7861}} & \cmark \\[0.5em]
    \rowcolor{gray!10} Claude 3.7 Sonnet (Thinking) & 45.9 ± 2.2 & \textbf{ \textcolor{darkgreen}{52.0 ± 2.2}} & 151.3s (2.5m) & \$1.0168 & \cmark \\[0.5em]
    Claude 3.7 Sonnet & 44.3 ± 2.2 & 49.5 ± 2.2 & 121.0s (2.0m) & \$0.9886 & \xmark \\[0.5em]
    \rowcolor{gray!10} o4 Mini & 37.3 ± 2.2 & 40.6 ± 2.1 & 164.4s (2.7m) & \$0.2863 & \cmark \\[0.5em]
    Grok 3 Mini Fast High Reasoning & 30.9 ± 1.9 & 36.3 ± 2.1 & 253.0s (4.2m) & \$0.1305 & \cmark \\[0.5em]
    \rowcolor{gray!10} Gemini 2.5 Pro Preview & 28.4 ± 2.0 & 33.0 ± 2.0 & 72.5s (1.2m) & \$0.1963 & \xmark \\[0.5em]
    GPT 4.1 & 26.7 ± 2.0 & 30.7 ± 2.0 & 61.3s (1.0m) & \$0.2309 & \xmark \\[0.5em]
    \rowcolor{gray!10} Grok 3 Beta & 25.8 ± 1.9 & 29.4 ± 2.0 & 60.5s (1.0m) & \$0.4254 & \xmark \\[0.5em]
    o1 & 21.4 ± 1.7 & 25.9 ± 1.9 & \textbf{ \textcolor{purple}{426.5s (7.1m)}} & \$1.4398 & \cmark \\[0.5em]
    \rowcolor{gray!10} GPT 4.1 mini & 20.3 ± 1.7 & 25.0 ± 1.9 & 51.6s (0.9m) & \$0.0684 & \xmark \\[0.5em]
    GPT 4o (2024-08-06) & 20.0 ± 1.7 & 24.6 ± 1.9 & 40.9s (0.7m) & \$0.2575 & \cmark \\[0.5em]
    \rowcolor{gray!10} Grok 3 Mini Fast Low Reasoning & 17.6 ± 1.6 & 21.4 ± 1.8 & 80.5s (1.3m) & \$0.0667 & \cmark \\[0.5em]
    Gemini 2.0 Flash (001) & 14.4 ± 1.4 & 18.2 ± 1.7 & 23.6s (0.4m) & \$0.0115 & \xmark \\[0.5em]
    \rowcolor{gray!10} Claude 3.5 Haiku Latest & 13.1 ± 1.4 & 17.1 ± 1.6 & 46.8s (0.8m) & \$0.0665 & \xmark \\[0.5em]
    o3 Mini & 12.8 ± 1.4 & 16.0 ± 1.6 & 146.1s (2.4m) & \$0.0472 & \cmark \\[0.5em]
    \rowcolor{gray!10} GPT 4o Mini & 10.8 ± 1.2 & 15.3 ± 1.6 & 93.4s (1.6m) & \$0.0375 & \xmark \\[0.5em]
    Mistral Small 3.1 (03/2025) & 10.8 ± 1.3 & 13.8 ± 1.5 & 39.1s (0.7m) & \$0.0061 & \xmark \\[0.5em]
    \rowcolor{gray!10} LLaMA 4 Scout & 5.8 ± 1.0 & 7.4 ± 1.1 & 13.8s (0.2m) & \$0.0046 & \xmark \\[0.5em]
    Command A & 4.6 ± 0.9 & 6.1 ± 1.0 & 94.9s (1.6m) & \$0.5628 & \xmark \\[0.5em]
    \rowcolor{gray!10} LLaMA 4 Maverick & 3.1 ± 0.8 & 3.7 ± 0.8 & 11.8s (0.2m) & \textbf{ \textcolor{darkgreen}{\$0.0023}} & \xmark \\[0.5em]
    LLaMA 3.3 Instruct Turbo (70B) & 2.8 ± 0.8 & 3.2 ± 0.8 & \textbf{ \textcolor{darkgreen}{3.3s (0.1m)}} & \$0.0030 & \xmark \\[0.5em]
    \rowcolor{gray!10} GPT 4.1 nano & \textbf{ \textcolor{purple}{2.4 ± 0.7}} & \textbf{ \textcolor{purple}{2.8 ± 0.7}} & 65.5s (1.1m) & \$0.0032 & \xmark \\
    \midrule\midrule
    \rowcolor{gray!10} Expert & - & - & \textbf{ \textcolor{purple}{1010.5s (16.8m)}} & \textbf{ \textcolor{purple}{\$25.66}} & - \\
    \bottomrule
    \end{tabular}
    }
    \caption{Results Table for all the models. Best values in bold green, worst in bold red.}
    \label{tab:full}
    \vspace{-10pt}
\end{table}

Figure \ref{fig:pareto_curve_subset} shows the subset of models that define the accuracy-cost Pareto curve (see the complete results in Figure \ref{fig:pareto_curve}). The models that define the Pareto curve and achieve over 20\% accuracy are o3, o4-mini, and GPT 4.1 mini. There is also a very clear logarithmic trend of cost versus accuracy, although some noticeable outliers include OpenAI's o3-mini and Cohere Command A, which fall significantly below the curve.

\subsubsection{Agent Tool Calling Analysis}
\label{section:agent_tool_calling_analysis}

\begin{figure}[t]
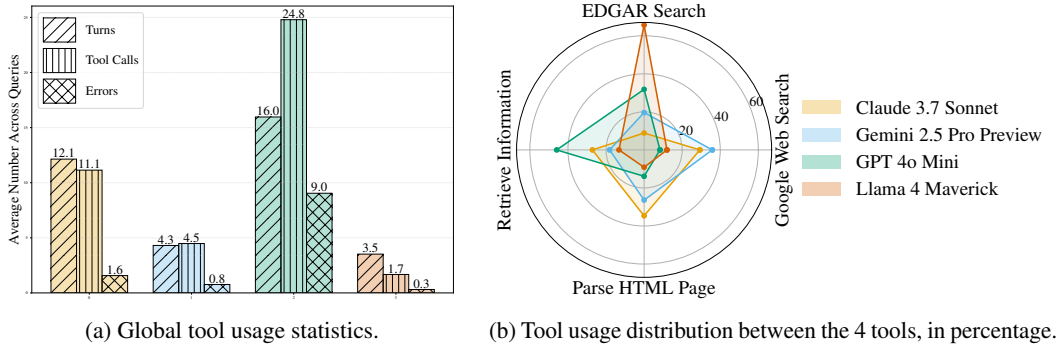

    \centering
    \begin{subfigure}[t]{0.43\textwidth}
        \centering
        \resizebox{\textwidth}{!}{%
            \input{figures/tikz/tool_calling_pattern.pgf}
        }
        \caption{Global tool usage statistics.}
        \label{fig:tool_calling_pattern}
    \end{subfigure}%
    \hfill
    \begin{subfigure}[t]{0.54\textwidth}
        \centering
        \resizebox{\textwidth}{!}{%
            \input{figures/tikz/spider_plot.pgf}
        }
        \caption{Tool usage distribution between the 4 tools, in percentage.}
        \label{fig:spider_graph}
    \end{subfigure}
    \caption{Overall tool use analysis. The best models not only better understand how to use the tools, they also tend to search deeper and be more persistent before settling on an answer. GPT 4o Mini is an outlier in this behavior, with a high number of unsuccessful tool calls.}
    \label{fig:tool_call_analysis_subfigures}
\end{figure}

Figure~\ref{fig:tool_call_analysis_subfigures} presents an overview of tool usage behavior across four representative models, chosen due to their markedly different strategies. A more exhaustive version covering all models is provided in the Appendix in Figure \ref{fig:tool_calling_all}. Subfigure~\ref{fig:tool_calling_pattern} shows global statistics, including the number of conversational turns and tool calls, as well as error rates. The number of turns taken to reach an answer ranges from 3.5 (LLaMA 4 Maverick) to 12.1 (Claude 3.7 Sonnet), while tool calls vary from 1.7 (LLaMA 4 Maverick) to an unusually high 24.8 (GPT-4o Mini).

\textbf{Insight 1: More exploratory models tend to perform better.} Claude 3.7 Sonnet, which ranks among the top performers, makes significantly more tool calls than either LLaMA 4 or Gemini, suggesting that more extensive exploration contributes to improved results. This behavior pattern also holds for the top performing model, o3 (see Figure \ref{fig:tool_calling_all}).

\textbf{Insight 2: High tool usage without precision leads to failure.} GPT-4o Mini stands out as an outlier, issuing a large number of tool calls, often in batches. However, this model suffers from the highest error rate by far, indicating poor tool utilization. It frequently fails by calling the same tool repeatedly (e.g., EDGAR Search) despite persistent errors, without adjusting its strategy.

\textbf{Insight 3: Balanced tool usage correlates with performance.} In Subfigure~\ref{fig:spider_graph}, which breaks down tool use across tools, higher-performing models such as Claude 3.7 Sonnet and Gemini 2.5 Pro Preview demonstrate balanced use across retrieval, search, and parsing operations. This indicates a nuanced understanding of when and how to apply specific tools.

\textbf{Insight 4: Tool misuse accounts for poor performance in some models.} LLaMA 4 Maverick, despite making fewer tool calls overall, shows disproportionate use of retrieval operations. In practice, this results in the model hallucinating documents and attempting to extract information from them, even when no such documents exist—highlighting a core misunderstanding of the tool interface.

\subsubsection{Case Study Examples}

\begin{figure}
  \centering
  \resizebox{\textwidth}{!}{
        \input{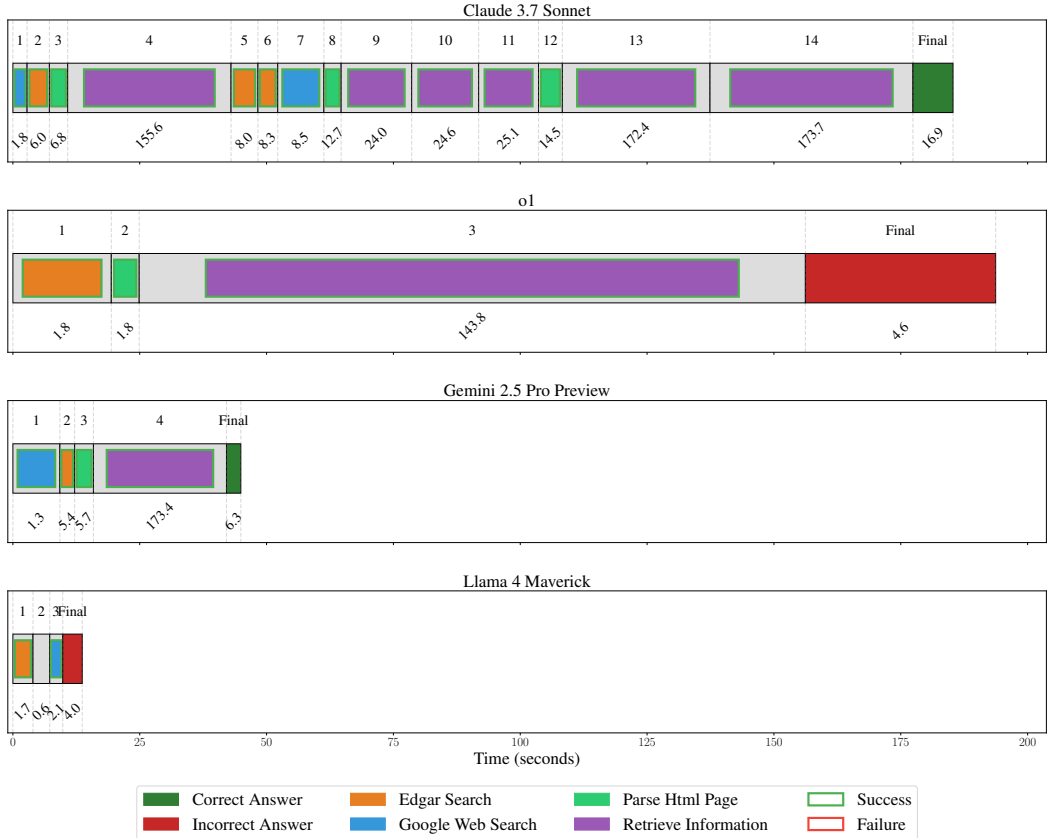}
    }
  \caption{Agent trajectories for various models. On each subplot, the numbers above the row are the turn index, and the numbers below are the number of tokens used for the turn (in thousands).}
  \label{fig:agent-trajectories}
\end{figure}

To highlight behavioral differences in tool usage, Figure~\ref{fig:agent-trajectories} presents agent trajectories for the task: \textit{Due to its business combinations, what is RTX Corp's (NYSE: RTX) projected future contractual obligation consumption for 2025 - 2029? Provide the amount for each year.} We visualize the interaction flow of Claude 3.7 Sonnet, o1, Gemini 2.5 Pro Preview, and LLaMA 4 Maverick, showing the sequence and timing of tool calls with outcomes.

As noted in Section~\ref{section:agent_tool_calling_analysis}, Claude 3.7 Sonnet explores tools extensively and efficiently. It follows a highly iterative strategy, repeatedly engaging with multiple tools even after successful calls. This verification approach aligns with its high tool call count and low error rate. Due to its speed, Claude 3.7 makes significantly more tool calls than o1 within the same 200-second timeframe, yet o1 fails to get the correct answer despite using the right tool sequence.

Gemini 2.5 Pro Preview displays the most streamlined trajectory, performing a minimal yet sufficient sequence to produce the correct result. While effective here, this tendency to use fewer tools reduces accuracy across the dataset.

LLaMA 4 Maverick misunderstands the purpose of the tools, calling EDGAR and Google searches but providing an answer without properly parsing and retrieving information from the results.

\section{Related Work}

\xhdr{General Autonomous-Agent Evaluation}  
LLMs have evolved from text predictors to autonomous agents \cite{cheng2024exploringlargelanguagemodel, xu2024theagentcompanybenchmarkingllmagents} capable of performing complex, multi-step tasks across domains. Modern agent frameworks enhance LLMs with access to external tools—such as calculators, search engines, and APIs—and allow them to reason, act, and decide when to stop autonomously~\cite{yao2023reactsynergizingreasoningacting, nakano2022webgptbrowserassistedquestionansweringhuman, schick2023toolformerlanguagemodelsteach}. This shift has spurred a wave of general-purpose agent benchmarks. 

AgentBench~\cite{liu2023agentbenchevaluatingllmsagents} evaluates LLM agents in interactive simulations spanning reasoning, web navigation, and games. {GTA (General Tool Agents)~\cite{wang2024gtabenchmarkgeneraltool} introduces real-user tool-augmented tasks covering web search, summarization, and booking. For coding tasks, SWE-Lancer~\cite{miserendino2025swelancerfrontierllmsearn} offers 1,488 real-world freelance problems sourced from Upwork. Web-based benchmarks such as GAIA \cite{mialon2023gaiabenchmarkgeneralai} and WebVoyager~\cite{he2024webvoyagerbuildingendtoendweb}  evaluate agentic browsing and user emulation across diverse sites.

Despite their diversity, these benchmarks rarely include fresh, time-sensitive questions. As a result, it's often unclear whether a model retrieved information during execution or memorized it during pretraining. Moreover, there is a lack of benchmarks targeting in-depth financial analysis—a critical domain where real-time, tool-augmented reasoning is essential. Our proposed FinanceAgent Benchmark addresses this gap with live EDGAR access, expert-written contemporary tasks, and strict grounding in evidence.

\xhdr{Finance-based NLP \& QA Datasets}  
Since their emergence into mainstream awareness, LLMs have been extensively investigated for their potential applications in financial services and analysis \cite{nie2024surveylargelanguagemodels}, to the point of training models for this sole purpose \cite{wu2023bloomberggptlargelanguagemodel}. Earlier work on finance QA centered on static datasets such as FinQA~\cite{chen2021finqa} and TAT-QA~\cite{zang2022tatqa}, which require numerical and hybrid reasoning over financial reports. These benchmarks contributed to specialized modeling approaches but did not support agentic planning or tool use.

Recent efforts have moved toward agent-style frameworks. \cite{zhang2024finqaagent} combines expert and critic LLMs for answer verification over structured filings, improving factual accuracy via multi-agent collaboration. FailSafeQA~\cite{vallath2024failsafeqa} evaluates robustness in finance QA with realistic, noisy queries based on EDGAR filings, but still assumes a static input and lacks autonomous planning. SWE-Lancer~\cite{miserendino2025swelancerfrontierllmsearn} is also relevant as a large-scale task benchmark, though it focuses on software engineering rather than finance.
FinAgent \cite{zhang2024multimodalfoundationagentfinancial} and MarS \cite{li2025marsfinancialmarketsimulation} explores the ability of an LLM-based agent to make decisions on the stock market based on multimodal observations: however, both focus on numerical data analysis rather than large documents.

In contrast, our FinanceAgent Benchmark evaluates autonomous agents on real-world financial workflows. It features 537 expert-authored, validated questions across nine categories, emphasizing multi-step reasoning, real-time retrieval from live SEC filings, and evidence-backed answers. Agents are equipped with tools like EDGAR search, calculators, and web access, making this the first benchmark to comprehensively test finance-specific agent capabilities in realistic, high-stakes environments.









\section{Conclusion}

Finance Agent Benchmark demonstrates significant limitations in current AI financial analysis capabilities. The best-performing model achieved only 46.8\% accuracy, highlighting a substantial gap between AI and human expert performance. The benchmark proved challenging across all tested systems, with performance varying dramatically between models (from below 3\% to 46.8\%).

Despite these limitations, all models completed tasks significantly faster than human experts, indicating potential productivity benefits even with current technological constraints. Performance followed a clear logarithmic cost-accuracy relationship, with specific models (o3, Claude 3.7 Sonnet, o4 Mini) defining the efficiency frontier. The stark performance differences across task types—with models handling simple retrieval better than complex financial modeling or market analysis—suggests that while AI agents show promise for automating routine financial tasks, considerable advancement is needed before these systems can reliably operate autonomously in high-stakes financial environments.

This benchmark provides a critical measure for tracking progress as foundation models continue to evolve toward meeting real-world financial analysis requirements. For future work, we recommend deeper investigation into model performance on structured tabular data or using more complex agents. Refer to detailed discussion in Appendix \ref{appendix:limitations}.

\begin{ack}
Thanks to the following people for their support: Alfston Thomas, Andrew Schettino, Kathy Ye, Kyle Jung, Matthew Friday, Michael Xia, and Nicholas Crawley-Brown.

This report was funded by Vals AI, a startup dedicated to evaluating Large Language Models

\end{ack}

{
\small
\bibliographystyle{plain}  
\bibliography{bibliography.bib}

\begin{thebibliography}{10}

\bibitem{an2024finverseautonomousagentversatile}
Siyu An, Qin Li, Junru Lu, Di~Yin, and Xing Sun.
\newblock Finverse: An autonomous agent system for versatile financial analysis, 2024.

\bibitem{bai2022constitutionalaiharmlessnessai}
Yuntao Bai, Saurav Kadavath, Sandipan Kundu, Amanda Askell, Jackson Kernion, Andy Jones, Anna Chen, Anna Goldie, Azalia Mirhoseini, Cameron McKinnon, Carol Chen, Catherine Olsson, Christopher Olah, Danny Hernandez, Dawn Drain, Deep Ganguli, Dustin Li, Eli Tran-Johnson, Ethan Perez, Jamie Kerr, Jared Mueller, Jeffrey Ladish, Joshua Landau, Kamal Ndousse, Kamile Lukosuite, Liane Lovitt, Michael Sellitto, Nelson Elhage, Nicholas Schiefer, Noemi Mercado, Nova DasSarma, Robert Lasenby, Robin Larson, Sam Ringer, Scott Johnston, Shauna Kravec, Sheer~El Showk, Stanislav Fort, Tamera Lanham, Timothy Telleen-Lawton, Tom Conerly, Tom Henighan, Tristan Hume, Samuel~R. Bowman, Zac Hatfield-Dodds, Ben Mann, Dario Amodei, Nicholas Joseph, Sam McCandlish, Tom Brown, and Jared Kaplan.
\newblock Constitutional ai: Harmlessness from ai feedback, 2022.

\bibitem{chen2021finqa}
Zichao Chen, Wenhu Chen, Yuwei Fang, Xiaocheng Feng, and Heng Ji.
\newblock Finqa: A dataset of numerical reasoning over financial data, 2021.

\bibitem{chen2025standardbenchmarks}
Zichen Chen, Jiaao Chen, Jianda Chen, and Misha Sra.
\newblock Position: Standard benchmarks fail -- llm agents present overlooked risks for financial applications, 2025.

\bibitem{cheng2024exploringlargelanguagemodel}
Yuheng Cheng, Ceyao Zhang, Zhengwen Zhang, Xiangrui Meng, Sirui Hong, Wenhao Li, Zihao Wang, Zekai Wang, Feng Yin, Junhua Zhao, and Xiuqiang He.
\newblock Exploring large language model based intelligent agents: Definitions, methods, and prospects, 2024.

\bibitem{bis2022fxturnover}
Bank for International~Settlements.
\newblock Triennial survey shows global foreign exchange trading averaged \$7.5 trillion a day in april 2022, 2022.

\bibitem{gao2023enablinglargelanguagemodels}
Tianyu Gao, Howard Yen, Jiatong Yu, and Danqi Chen.
\newblock Enabling large language models to generate text with citations, 2023.

\bibitem{he2024webvoyagerbuildingendtoendweb}
Hongliang He, Wenlin Yao, Kaixin Ma, Wenhao Yu, Yong Dai, Hongming Zhang, Zhenzhong Lan, and Dong Yu.
\newblock Webvoyager: Building an end-to-end web agent with large multimodal models, 2024.

\bibitem{kapoor2024aiagents}
Sayash Kapoor, Benedikt Stroebl, Zachary~S. Siegel, Nitya Nadgir, and Arvind Narayanan.
\newblock Ai agents that matter, 2024.

\bibitem{li2025marsfinancialmarketsimulation}
Junjie Li, Yang Liu, Weiqing Liu, Shikai Fang, Lewen Wang, Chang Xu, and Jiang Bian.
\newblock Mars: a financial market simulation engine powered by generative foundation model, 2025.

\bibitem{liu2023rethinkingtabulardataunderstanding}
Tianyang Liu, Fei Wang, and Muhao Chen.
\newblock Rethinking tabular data understanding with large language models, 2023.

\bibitem{liu2023agentbenchevaluatingllmsagents}
Xiao Liu, Hao Yu, Hanchen Zhang, Yifan Xu, Xuanyu Lei, Hanyu Lai, Yu~Gu, Hangliang Ding, Kaiwen Men, Kejuan Yang, Shudan Zhang, Xiang Deng, Aohan Zeng, Zhengxiao Du, Chenhui Zhang, Sheng Shen, Tianjun Zhang, Yu~Su, Huan Sun, Minlie Huang, Yuxiao Dong, and Jie Tang.
\newblock Agentbench: Evaluating llms as agents, 2023.

\bibitem{liu2023GEvalNE}
Yang Liu, Dan Iter, Yichong Xu, Shuo Wang, Ruochen Xu, and Chenguang Zhu.
\newblock G-eval: Nlg evaluation using gpt-4 with better human alignment.
\newblock In {\em Conference on Empirical Methods in Natural Language Processing}, 2023.

\bibitem{mialon2023gaiabenchmarkgeneralai}
Grégoire Mialon, Clémentine Fourrier, Craig Swift, Thomas Wolf, Yann LeCun, and Thomas Scialom.
\newblock Gaia: a benchmark for general ai assistants, 2023.

\bibitem{miserendino2025swelancerfrontierllmsearn}
Samuel Miserendino, Michele Wang, Tejal Patwardhan, and Johannes Heidecke.
\newblock Swe-lancer: Can frontier llms earn \$1 million from real-world freelance software engineering?, 2025.

\bibitem{nakano2022webgptbrowserassistedquestionansweringhuman}
Reiichiro Nakano, Jacob Hilton, Suchir Balaji, Jeff Wu, Long Ouyang, Christina Kim, Christopher Hesse, Shantanu Jain, Vineet Kosaraju, William Saunders, Xu~Jiang, Karl Cobbe, Tyna Eloundou, Gretchen Krueger, Kevin Button, Matthew Knight, Benjamin Chess, and John Schulman.
\newblock Webgpt: Browser-assisted question-answering with human feedback, 2022.

\bibitem{nie2024surveylargelanguagemodels}
Yuqi Nie, Yaxuan Kong, Xiaowen Dong, John~M. Mulvey, H.~Vincent Poor, Qingsong Wen, and Stefan Zohren.
\newblock A survey of large language models for financial applications: Progress, prospects and challenges, 2024.

\bibitem{Ouyang_2025}
Shuyin Ouyang, Jie~M. Zhang, Mark Harman, and Meng Wang.
\newblock An empirical study of the non-determinism of chatgpt in code generation.
\newblock {\em ACM Transactions on Software Engineering and Methodology}, 34(2):1–28, January 2025.

\bibitem{pathak2025rubricneedenhancingllmbased}
Aditya Pathak, Rachit Gandhi, Vaibhav Uttam, Devansh, Yashwanth Nakka, Aaryan~Raj Jindal, Pratyush Ghosh, Arnav Ramamoorthy, Shreyash Verma, Aditya Mittal, Aashna Ased, Chirag Khatri, Jagat~Sesh Challa, and Dhruv Kumar.
\newblock Rubric is all you need: Enhancing llm-based code evaluation with question-specific rubrics, 2025.

\bibitem{vallath2024failsafeqa}
Rahul~Vallath Prabhakar, Jiaqi Zhou, Zichen Chen, and Misha Sra.
\newblock Failsafeqa: Evaluating the reliability of llms on financial tasks, 2024.

\bibitem{pwc_effectiveness_benchmark}
PWC.
\newblock Stepping up how finance functions are transforming to drive business results, 2017.

\bibitem{schick2023toolformerlanguagemodelsteach}
Timo Schick, Jane Dwivedi-Yu, Roberto Dessì, Roberta Raileanu, Maria Lomeli, Luke Zettlemoyer, Nicola Cancedda, and Thomas Scialom.
\newblock Toolformer: Language models can teach themselves to use tools, 2023.

\bibitem{serp_api}
serpapi.com.
\newblock Serpapi.

\bibitem{starace2025paperbenchevaluatingaisability}
Giulio Starace, Oliver Jaffe, Dane Sherburn, James Aung, Jun~Shern Chan, Leon Maksin, Rachel Dias, Evan Mays, Benjamin Kinsella, Wyatt Thompson, Johannes Heidecke, Amelia Glaese, and Tejal Patwardhan.
\newblock Paperbench: Evaluating ai's ability to replicate ai research, 2025.

\bibitem{tang2024strucbenchlargelanguagemodels}
Xiangru Tang, Yiming Zong, Jason Phang, Yilun Zhao, Wangchunshu Zhou, Arman Cohan, and Mark Gerstein.
\newblock Struc-bench: Are large language models really good at generating complex structured data?, 2024.

\bibitem{thakur2025judgingjudgesevaluatingalignment}
Aman~Singh Thakur, Kartik Choudhary, Venkat~Srinik Ramayapally, Sankaran Vaidyanathan, and Dieuwke Hupkes.
\newblock Judging the judges: Evaluating alignment and vulnerabilities in llms-as-judges, 2025.

\bibitem{wang2024gtabenchmarkgeneraltool}
Jize Wang, Zerun Ma, Yining Li, Songyang Zhang, Cailian Chen, Kai Chen, and Xinyi Le.
\newblock Gta: A benchmark for general tool agents, 2024.

\bibitem{williamson2024hallucinations}
Eric Williamson.
\newblock Understanding ai hallucinations: What's the problem and how can we address it?, 2024.

\bibitem{wu2023bloomberggptlargelanguagemodel}
Shijie Wu, Ozan Irsoy, Steven Lu, Vadim Dabravolski, Mark Dredze, Sebastian Gehrmann, Prabhanjan Kambadur, David Rosenberg, and Gideon Mann.
\newblock wu2023bloomberggptlargelanguagemodel, 2023.

\bibitem{xu2024benchmarkdatacontaminationlarge}
Cheng Xu, Shuhao Guan, Derek Greene, and M-Tahar Kechadi.
\newblock Benchmark data contamination of large language models: A survey, 2024.

\bibitem{xu2024theagentcompanybenchmarkingllmagents}
Frank~F. Xu, Yufan Song, Boxuan Li, Yuxuan Tang, Kritanjali Jain, Mengxue Bao, Zora~Z. Wang, Xuhui Zhou, Zhitong Guo, Murong Cao, Mingyang Yang, Hao~Yang Lu, Amaad Martin, Zhe Su, Leander Maben, Raj Mehta, Wayne Chi, Lawrence Jang, Yiqing Xie, Shuyan Zhou, and Graham Neubig.
\newblock Theagentcompany: Benchmarking llm agents on consequential real world tasks, 2024.

\bibitem{yao2023reactsynergizingreasoningacting}
Shunyu Yao, Jeffrey Zhao, Dian Yu, Nan Du, Izhak Shafran, Karthik Narasimhan, and Yuan Cao.
\newblock React: Synergizing reasoning and acting in language models, 2023.

\bibitem{yehudai2025surveyevaluationllmbasedagents}
Asaf Yehudai, Lilach Eden, Alan Li, Guy Uziel, Yilun Zhao, Roy Bar-Haim, Arman Cohan, and Michal Shmueli-Scheuer.
\newblock Survey on evaluation of llm-based agents, 2025.

\bibitem{zang2022tatqa}
Xinyu Zang, Zichao Li, Wenhao Yu, Xiaodong Liu, Ivan Evtimov, Muhao Chen, Yejin Choi, and Hannaneh Hajishirzi.
\newblock Tat-qa: A question answering benchmark on a hybrid of tabular and textual content in finance.
\newblock In {\em Proceedings of ACL}, 2022.

\bibitem{zhang2024finqaagent}
Bo~Zhang, Wenhao Yu, Weixin Liang, and Wenhu Chen.
\newblock Enhancing financial question answering with a multi-agent reflection framework, 2024.

\bibitem{zhang2024multimodalfoundationagentfinancial}
Wentao Zhang, Lingxuan Zhao, Haochong Xia, Shuo Sun, Jiaze Sun, Molei Qin, Xinyi Li, Yuqing Zhao, Yilei Zhao, Xinyu Cai, Longtao Zheng, Xinrun Wang, and Bo~An.
\newblock A multimodal foundation agent for financial trading: Tool-augmented, diversified, and generalist, 2024.

\end{thebibliography}
}

\newpage


\begin{appendices}

\section{Dataset Creation}

\begin{table}[htbp]
    \centering
    \renewcommand{\arraystretch}{1.1}
    \begin{tabular}{p{0.22\textwidth}p{0.45\textwidth}p{0.15\textwidth}p{0.08\textwidth}}
    \toprule
    \textbf{Task Name} & \textbf{Description} & \textbf{Difficulty Level} & \textbf{Count} \\
    \midrule
    Quantitative Retrieval & Direct extraction of numerical information from one or more documents without any post-retrieval calculation or manipulation. & Easy & 102 (19\%) \\
    \rowcolor[gray]{0.95}
    Qualitative Retrieval & Direct quotation or summarization of non-numerical information from one or more documents. & Easy & 97 (97\%) \\
    Numerical Reasoning & Calculations or aggregation of key numbers to produce an answer. & Easy & 83 (15\%) \\
    \rowcolor[gray]{0.95}
    Complex Retrieval & Numerical or non-numerical retrieval or content summarization requiring synthesis of information from multiple documents. & Medium & 29 (6\%) \\
    Adjustments & Quantitative and qualitative analysis of reporting context bridging GAAP and Non-GAAP Financial Metrics. & Medium & 43 (8\%)\\
    \rowcolor[gray]{0.95}
    Beat or Miss & Comparison of forward management guidance versus actuals, synthesized by reconciling sequential quarterly reporting documents. & Medium & 69 (13\%)\\
    Trends & Analyze patterns within a single company's reporting structure or calculate and contextualize evolving performance, key metrics or business composition. & Hard & 33 (6\%) \\
    \rowcolor[gray]{0.95}
    Financial Modeling & Complex numerical reasoning calculations which require additional financial expertise to define and evaluate. & Hard & 47 (9\%) \\
    Market Analysis & Advanced analysis of one or more companies using various documents, requiring normalization of comparison metrics, or complex reasoning and usage of causality to contextualize drivers of business changes or competition dynamics. & Hard & 34 (6\%)\\
    \bottomrule
    \end{tabular}
    \caption{Task Types, Descriptions, Analyst Difficulty Levels and Distribution}
    \label{table:taxonomy-description}
\end{table}

\subsection{Rubric Generation Prompts}

To generate the rubrics for the questions and before the final review, we used GPT-4o with the following prompts that make use of the \textit{Questions}, \textit{Reasoning Steps}, and \textit{Answers}:

\begin{tcolorbox}[title=System Prompt,fonttitle=\bfseries]
You are an expert at converting answers into structured evaluation checks.\\
Your task is to analyze an answer and break it down into individual checks that can be automatically evaluated.\\

Each check must use one of these available operators:\\
\quad - edgar\_research\_operator: This operator checks that the given criteria is present in the text as a complete, meaningful concept. It verifies factual content such as numerical figures (accepting rounded values), names, dates, and relationships between facts. Each check should represent a complete piece of information rather than fragmenting related facts into separate checks. The format, writing style, and length of the answer do not affect this check.\\

Guidelines:\\
1. Break down complex answers into multiple simple checks only when the answer contains distinct, separable components.\\
2. Create checks that are specific, measurable, and objective.\\
3. Ensure each criteria is clear, precise, and unambiguous. Do not write full sentences for the criteria if not necessary. It can just be figures or phrases.\\
\end{tcolorbox}

\begin{tcolorbox}[title=User Prompt,fonttitle=\bfseries]
Convert this answer into a list of specific evaluation checks.\\
Break down complex requirements into multiple simple checks where appropriate.\\
Return the result as a JSON array of objects with 'operator' and 'criteria' fields.\\

Important: The question and reasoning are provided ONLY for context to help you understand the answer. \\
Your checks must ONLY evaluate the answer itself - not the question or reasoning.\\
The checks will be applied exclusively to the answer text.\\

Create meaningful checks that capture substantive elements of the answer. Each check should:\\
- Evaluate a significant aspect of the answer\\
- Be clearly defined and testable\\
- Make sense as a standalone evaluation criterion\\

Very important:\\
- Do not split sentences or phrases into multiple checks if they are related to the same underlying concept.\\
- Some answers might be a bit verbose and contain more information than originally asked in the question. Do not make more checks than what is asked in the question. For example, if a question asks about a specific number, and the answer contains the number but also the calculation, do not make two checks: one for the number and one for the calculation. Make only one check for the number.\\
- Particularly make sure that the logical connections are kept within the same check and not split into multiple checks.\\

Data to evaluate:\\
Question: \{question\}\\
Reasoning: \{reasoning\}\\
Answer: \{answer\}
\end{tcolorbox}

\subsection{Evaluation Example}
\label{appendix:evaluation-example}
To illustrate our approach, consider the following question from our dataset:

\textit{``How has Carvana's (NYSE: CVNA) gross profit per unit changed from 2019-2024?''}

The expert's answer contains several key facts:

\textit{``CVNA has increased its gross profit per unit from \$2,852 in 2019 to \$7,196 in 2024 representing a 20.3\% CAGR. It has driven this metric through improved operating and technology efficiencies as well as introducing new products such as origination fees from financing.''}

Instead of evaluating this answer as a whole, we decompose it into five distinct checks:
\begin{enumerate}[leftmargin=*,itemsep=0pt,parsep=0pt,topsep=0pt]
    \item CVNA gross profit per unit in 2019 was \$2,852
    \item CVNA gross profit per unit in 2024 was \$7,196
    \item CVNA gross profit per unit increased at a 20.3\% CAGR from 2019 to 2024
    \item Increase driven by improved operating and technology efficiencies
    \item Increase driven by new products such as origination fees from financing
\end{enumerate}

Each point is evaluated separately using a dedicated evaluation check. Additionally, we employ a contradiction check, against the entire expert answer to ensure the generated response doesn't contain any conflicting information. 

\section{Dataset Access}
\label{appendix:dataset-split}

\subsection{Accessibility and Reproducibility}
\label{appendix:reproducibility}
The dataset and code are released under permissive open-source licenses: CC BY 4.0 for the dataset and MIT License for the code. Users are free to use and adapt the resources, provided they include appropriate citation. A DOI is provided for the dataset on HuggingFace: \url{https://doi.org/10.57967/hf/5514}. A different DOI is provided via Zenodo for the harness: \url{https://doi.org/10.5281/zenodo.15428823}.

The public validation set and agent code are open source and accessible at: \url{https://github.com/vals-ai/finance-agent}. This repository includes the dataset, simulation environment, evaluation scripts, and detailed documentation to support reproducibility and further research.

The dataset is provided in standard CSV format, accompanied by a structured README and usage examples. The simulation environment is implemented in Python using standard libraries, with clear setup instructions.

We provide documentation based on the Datasheets for Datasets framework, outlining data collection, intended uses, and limitations. All data used is sourced from public, license-compatible domains. The authors bear full responsibility for the dataset and confirm compliance with all applicable rights and licensing.

To ensure long-term accessibility, the dataset is hosted on both GitHub and Zenodo. Structured metadata is included via Zenodo to support discoverability. All experiments in the paper can be reproduced using the provided code and instructions.

\subsection{Dataset Split Correlation Plots}

The datasets were split to maintain a consistent distribution of question categories across the training, validation, and test sets. In Figure \ref{fig:correlation-plot-unbalanced}, we plot the accuracies of various models on the Validation set (Public + Private) against their corresponding accuracies on the Test set, with each point representing one model. We observe a strong linear relationship, achieving a Pearson correlation coefficient of $0.98$ and an $R^2$ value of $0.97$.

The experiments and results described in this document were conducted on all 537 samples, as no data had been made public at the time of experimentation. We maintain a public leaderboard showing model performance on \textbf{the test set only} at \href{https://www.vals.ai/benchmarks/finance\_agent}{vals.ai}.

\begin{figure}
    \centering
    \input{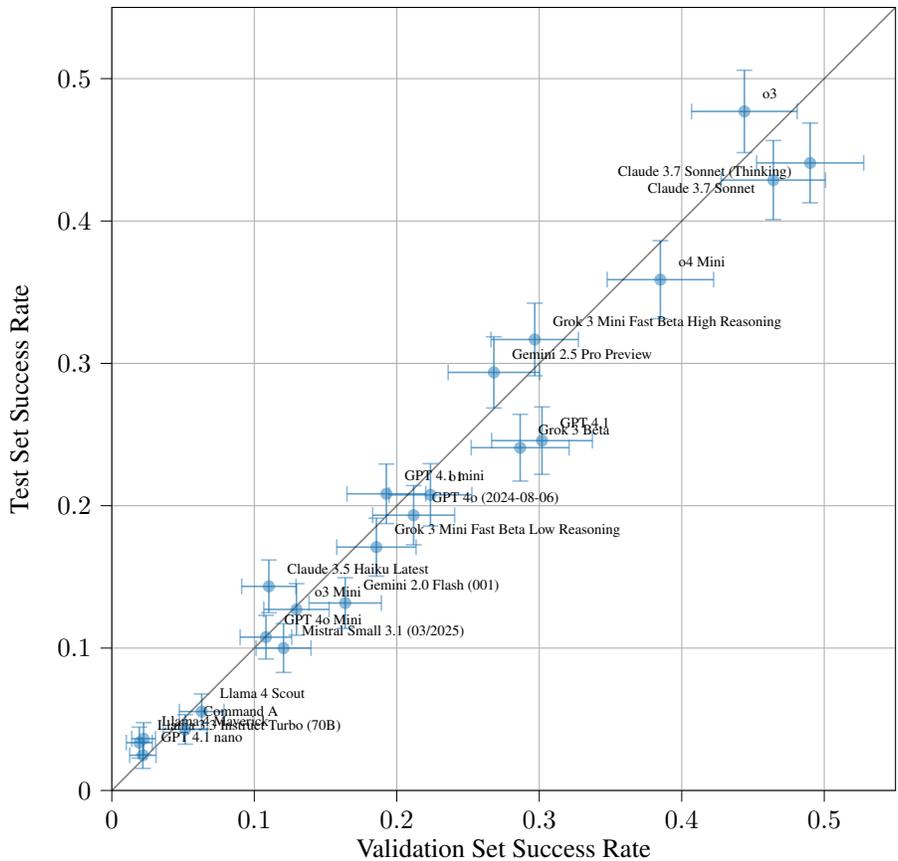}
    \caption{Correlation Plot between Validation and Test sets for Class-Balanced accuracy.}
    \label{fig:correlation-plot-unbalanced}
\end{figure}

\section{Harness Details}

\subsection{Instruction Prompt}
\label{appendix:instruction-prompt}
The models were all prompted with the following instructions, as well as the list of tools available. We provide the current date to models in the prompt, because otherwise they tend to guess the date, and are not always correct. 

\begin{tcolorbox}[title=Instruction Prompt,fonttitle=\bfseries]
You are a financial agent. Today is \{current\_date\}. You are given a question 
and you need to answer it using the tools provided.
You may not interact with the user.\\
When you have the answer, you should respond with 'FINAL ANSWER:' followed by 
your answer.\\
At the end of your answer, you should provide your sources in a dictionary with 
the following format:
\begin{alltt}
\{
  "sources": [
      \{
          "url": "https://example.com",
          "name": "Name of the source"
      \},
      ...
  ]
\}
\end{alltt}
Question:\\
\{question\}
\end{tcolorbox}

\subsection{Tool Use}
\label{appendix:tool-use}

\subsubsection{Google Web Search}
A tool that leverages SerpAPI \cite{serp_api} to perform web searches.

\textbf{Input Arguments:}
\begin{itemize}
    \item \texttt{search\_query} (string, required): The query to search for information
\end{itemize}

\textbf{Implementation Details:}
\begin{itemize}
    \item Returns up to 10 results by default
    \item Uses SerpAPI to access Google Search results
    \item Results include title, link, and snippet information
\end{itemize}

\subsubsection{EDGAR Search}
A tool for searching the SEC's EDGAR database using the SEC API.

\textbf{Input Arguments:}
\begin{itemize}
    \item \texttt{query} (string, required): Keywords or phrases to search (e.g., "substantial doubt" OR "material weakness")
    \item \texttt{form\_types} (array, required): SEC form types to limit search to (e.g., ["8-K", "10-Q"])
    \item \texttt{ciks} (array, required): Company CIK numbers to filter results
    \item \texttt{start\_date} (string, required): Start date in yyyy-mm-dd format
    \item \texttt{end\_date} (string, required): End date in yyyy-mm-dd format
    \item \texttt{page} (string, required): Pagination parameter
    \item \texttt{top\_n\_results} (integer, required): Number of top results to return
\end{itemize}

\textbf{Implementation Details:}
\begin{itemize}
    \item Returns filing metadata, not the full filing text
    \item Optimized for financial document discovery
\end{itemize}

\subsubsection{Parse HTML Page}
A tool that extracts and saves content from web pages.

\textbf{Input Arguments:}
\begin{itemize}
    \item \texttt{url} (string, required): The URL of the HTML page to parse
    \item \texttt{key} (string, required): The key to use when saving the result in the data structure
\end{itemize}

\textbf{Implementation Details:}
\begin{itemize}
    \item Uses BeautifulSoup to parse HTML content
    \item Extracts clean text by removing scripts, styles, and formatting
    \item Saves content to a key-value store store later retrieval
\end{itemize}

\subsubsection{Retrieve Information}
A tool that retrieves stored document content and processes it through the LLM.

\textbf{Input Arguments:}
\begin{itemize}
    \item \texttt{prompt} (string, required): The prompt containing placeholder(s) in format \{\{key\_name\}\}
    \item \texttt{input\_character\_ranges} (object, optional): Dictionary mapping keys to character ranges [start, end]
\end{itemize}

\textbf{Implementation Details:}
\begin{itemize}
    \item Replaces \{\{key\_name\}\} placeholders with actual document content
    \item Allows extracting specific portions of documents using character ranges
    \item Makes a call to the same LLM that's leading the agent to process the information
\end{itemize}

\subsection{Error Handling}
\label{appendix:error-handling}

\subsubsection{Rate Limit Errors}
\begin{itemize}
    \item Implements exponential backoff with randomized jitter for 429 errors
    \item Automatically retries API calls when rate limits are encountered
    \item Maximum of 8 retry attempts with increasing delays
\end{itemize}

\subsubsection{Token Limit Errors}
\begin{itemize}
    \item \textbf{During retrieval}: Errors from documents that are too large are returned to the agent, which can retry by sending partial document chunks
    \item \textbf{During conversation}: If the conversation exceeds the token limit, older messages are removed until the entire exchange fits within the allowed limits. We chose not to implement a more sophisticated long-short term memory mechanism, as the model rarely exhausted its context window during typical conversations. In most instances where this did occur, it was due to the model repeatedly attempting—unsuccessfully—to use the same tool, a behavior we categorize as a model error mode.
\end{itemize}

\subsubsection{Argument Formatting Errors}
\begin{itemize}
    \item Input validation errors are caught and returned to the agent
    \item Detailed error messages help the agent understand and correct formatting issues
    \item Handles special cases like JSON string parsing for arrays
\end{itemize}

\section{Limitations}
\label{appendix:limitations}
\subsection{Dataset}

Our current dataset emphasizes questions that can be answered with relatively short passages or that focus on the final output of a more complex reasoning process. During development, domain experts highlighted that a significant portion of their workflows involves interacting with structured tabular data, such as spreadsheets or CSV files. While we included a few examples of such queries, future work could more deeply investigate model performance in synthesizing or reasoning over entire tabular documents and financial datasets. Prior research has begun exploring this direction, demonstrating the challenges and opportunities of reasoning over structured financial data \cite{tang2024strucbenchlargelanguagemodels, liu2023rethinkingtabulardataunderstanding}.

Moreover, although we asked models to provide citations or sources for their answers, we did not evaluate the accuracy or reliability of these sources. This is an important avenue for future work, especially in high-stakes domains. Recent work on source attribution and citation accuracy in language models could guide such evaluation \cite{gao2023enablinglargelanguagemodels}.

\subsection{Agentic Harness}

The agentic harness used in our experiments follows a relatively simple architecture inspired by the ReAct framework, enabling models to interleave reasoning and action \cite{yao2023reactsynergizingreasoningacting}. Our setup is comparable in spirit to benchmarks such as PaperBench \cite{starace2025paperbenchevaluatingaisability} and SWE-Lancer \cite{miserendino2025swelancerfrontierllmsearn}. While sufficient to evaluate general capabilities, it is not representative of the full complexity of commercial retrieval-augmented generation (RAG) systems or proprietary agentic tools developed by industry leaders such as OpenAI, Google, or xAI.

Future research should explore how these more advanced systems—many of which integrate proprietary retrieval mechanisms, structured data backends, or user feedback loops—perform on similar document understanding tasks. Additionally, our models only had access to publicly available SEC filings and were not integrated with private knowledge bases or advanced retrieval systems. Enabling access to broader and deeper data sources, including internal contracts and proprietary databases, could significantly impact performance in real-world use cases.

\section{Additional Analysis}

\subsection{General Analysis}

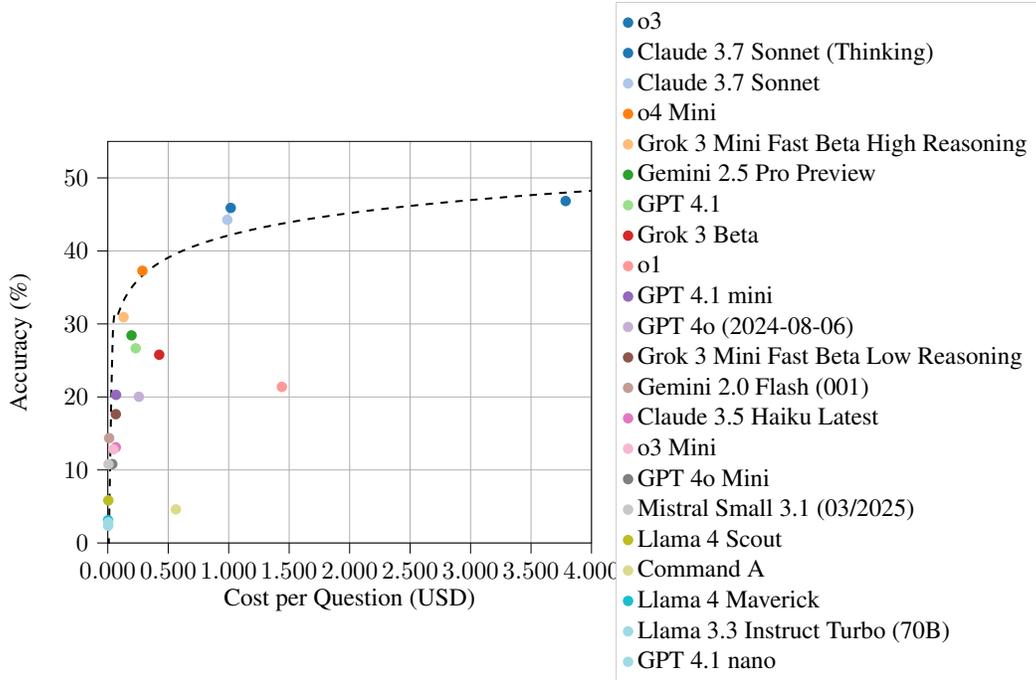
\begin{figure}[h]
    \begin{centering}
    \resizebox{\textwidth}{!}{%
\begin{tikzpicture}

\definecolor{crimson2143940}{RGB}{214,39,40}
\definecolor{darkgray176}{RGB}{176,176,176}
\definecolor{darkorange25512714}{RGB}{255,127,14}
\definecolor{darkturquoise23190207}{RGB}{23,190,207}
\definecolor{forestgreen4416044}{RGB}{44,160,44}
\definecolor{goldenrod18818934}{RGB}{188,189,34}
\definecolor{gray127}{RGB}{127,127,127}
\definecolor{khaki219219141}{RGB}{219,219,141}
\definecolor{lightblue158218229}{RGB}{158,218,229}
\definecolor{lightgray204}{RGB}{204,204,204}
\definecolor{lightgreen152223138}{RGB}{152,223,138}
\definecolor{lightsalmon255152150}{RGB}{255,152,150}
\definecolor{lightsalmon255187120}{RGB}{255,187,120}
\definecolor{lightsteelblue174199232}{RGB}{174,199,232}
\definecolor{mediumpurple148103189}{RGB}{148,103,189}
\definecolor{orchid227119194}{RGB}{227,119,194}
\definecolor{pink247182210}{RGB}{247,182,210}
\definecolor{rosybrown196156148}{RGB}{196,156,148}
\definecolor{sienna1408675}{RGB}{140,86,75}
\definecolor{silver199}{RGB}{199,199,199}
\definecolor{steelblue31119180}{RGB}{31,119,180}
\definecolor{thistle197176213}{RGB}{197,176,213}

\begin{axis}[
legend cell align={left},
legend style={
  fill opacity=1,
  draw opacity=1,
  text opacity=1,
  at={(1.05,0.5)},
  anchor=west,
  draw=lightgray204
},
tick align=outside,
tick pos=left,
x grid style={darkgray176},
xlabel={Cost per Question (USD)},
xmajorgrids,
xmin=0, xmax=4,
xtick style={color=black},
xtick={0,0.5,1,1.5,2,2.5,3,3.5,4},
xticklabels={
  \(\displaystyle 0.000\),
  \(\displaystyle 0.500\),
  \(\displaystyle 1.000\),
  \(\displaystyle 1.500\),
  \(\displaystyle 2.000\),
  \(\displaystyle 2.500\),
  \(\displaystyle 3.000\),
  \(\displaystyle 3.500\),
  \(\displaystyle 4.000\)
},
y grid style={darkgray176},
ylabel={Accuracy (\%)},
ymajorgrids,
ymin=0, ymax=55,
ytick style={color=black}
]
\addplot [semithick, steelblue31119180, opacity=1, mark=*, mark size=1.8, mark options={solid}, only marks]
table {%
3.78610164342291 46.837344874371
};
\addlegendentry{o3}
\addplot [semithick, steelblue31119180, opacity=1, mark=*, mark size=1.8, mark options={solid}, only marks]
table {%
1.01676200923806 45.8904948865151
};
\addlegendentry{Claude 3.7 Sonnet (Thinking)}
\addplot [semithick, lightsteelblue174199232, opacity=1, mark=*, mark size=1.8, mark options={solid}, only marks]
table {%
0.988625833099283 44.2639091676131
};
\addlegendentry{Claude 3.7 Sonnet}
\addplot [semithick, darkorange25512714, opacity=1, mark=*, mark size=1.8, mark options={solid}, only marks]
table {%
0.2862863155701 37.276820487755
};
\addlegendentry{o4 Mini}
\addplot [semithick, lightsalmon255187120, opacity=1, mark=*, mark size=1.8, mark options={solid}, only marks]
table {%
0.13049179427344 30.9238811654442
};
\addlegendentry{Grok 3 Mini Fast Beta High Reasoning}
\addplot [semithick, forestgreen4416044, opacity=1, mark=*, mark size=1.8, mark options={solid}, only marks]
table {%
0.196306608510841 28.433138978131
};
\addlegendentry{Gemini 2.5 Pro Preview}
\addplot [semithick, lightgreen152223138, opacity=1, mark=*, mark size=1.8, mark options={solid}, only marks]
table {%
0.230873824907237 26.6630122657041
};
\addlegendentry{GPT 4.1}
\addplot [semithick, crimson2143940, opacity=1, mark=*, mark size=1.8, mark options={solid}, only marks]
table {%
0.425365855975696 25.782982697717
};
\addlegendentry{Grok 3 Beta}
\addplot [semithick, lightsalmon255152150, opacity=1, mark=*, mark size=1.8, mark options={solid}, only marks]
table {%
1.43975413124239 21.3825657020688
};
\addlegendentry{o1}
\addplot [semithick, mediumpurple148103189, opacity=1, mark=*, mark size=1.8, mark options={solid}, only marks]
table {%
0.0683812668820963 20.2872449332544
};
\addlegendentry{GPT 4.1 mini}
\addplot [semithick, thistle197176213, opacity=1, mark=*, mark size=1.8, mark options={solid}, only marks]
table {%
0.257491518299961 20.0236667533797
};
\addlegendentry{GPT 4o (2024-08-06)}
\addplot [semithick, sienna1408675, opacity=1, mark=*, mark size=1.8, mark options={solid}, only marks]
table {%
0.0666693135255327 17.6378660621282
};
\addlegendentry{Grok 3 Mini Fast Beta Low Reasoning}
\addplot [semithick, rosybrown196156148, opacity=1, mark=*, mark size=1.8, mark options={solid}, only marks]
table {%
0.0114746008157894 14.3555320942525
};
\addlegendentry{Gemini 2.0 Flash (001)}
\addplot [semithick, orchid227119194, opacity=1, mark=*, mark size=1.8, mark options={solid}, only marks]
table {%
0.0664669898632701 13.1040857457352
};
\addlegendentry{Claude 3.5 Haiku Latest}
\addplot [semithick, pink247182210, opacity=1, mark=*, mark size=1.8, mark options={solid}, only marks]
table {%
0.0471560993470098 12.8354487361899
};
\addlegendentry{o3 Mini}
\addplot [semithick, gray127, opacity=1, mark=*, mark size=1.8, mark options={solid}, only marks]
table {%
0.0374871363283261 10.7992263915581
};
\addlegendentry{GPT 4o Mini}
\addplot [semithick, silver199, opacity=1, mark=*, mark size=1.8, mark options={solid}, only marks]
table {%
0.00605787514245645 10.773305418569
};
\addlegendentry{Mistral Small 3.1 (03/2025)}
\addplot [semithick, goldenrod18818934, opacity=1, mark=*, mark size=1.8, mark options={solid}, only marks]
table {%
0.00459797897284824 5.82624345610504
};
\addlegendentry{Llama 4 Scout}
\addplot [semithick, khaki219219141, opacity=1, mark=*, mark size=1.8, mark options={solid}, only marks]
table {%
0.562849864674662 4.60100359128803
};
\addlegendentry{Command A}
\addplot [semithick, darkturquoise23190207, opacity=1, mark=*, mark size=1.8, mark options={solid}, only marks]
table {%
0.00233937247067305 3.11206235775217
};
\addlegendentry{Llama 4 Maverick}
\addplot [semithick, lightblue158218229, opacity=1, mark=*, mark size=1.8, mark options={solid}, only marks]
table {%
0.00302138270884077 2.82381188702667
};
\addlegendentry{Llama 3.3 Instruct Turbo (70B)}
\addplot [semithick, lightblue158218229, opacity=1, mark=*, mark size=1.8, mark options={solid}, only marks]
table {%
0.00318040528592362 2.36047565074661
};
\addlegendentry{GPT 4.1 nano}

\addplot [
    thick,
    dashed,
    domain=0.00001:4,
    samples=100,
    smooth
]
{4.4*ln(x)+42.13};

\end{axis}
\end{tikzpicture}
    }
    \caption{Cost-Accuracy pareto curve results on \ours{}. }
    \label{fig:pareto_curve}
    \end{centering}
\end{figure}

\begin{figure}[h]
    \centering
    \resizebox{\textwidth}{!}{\input{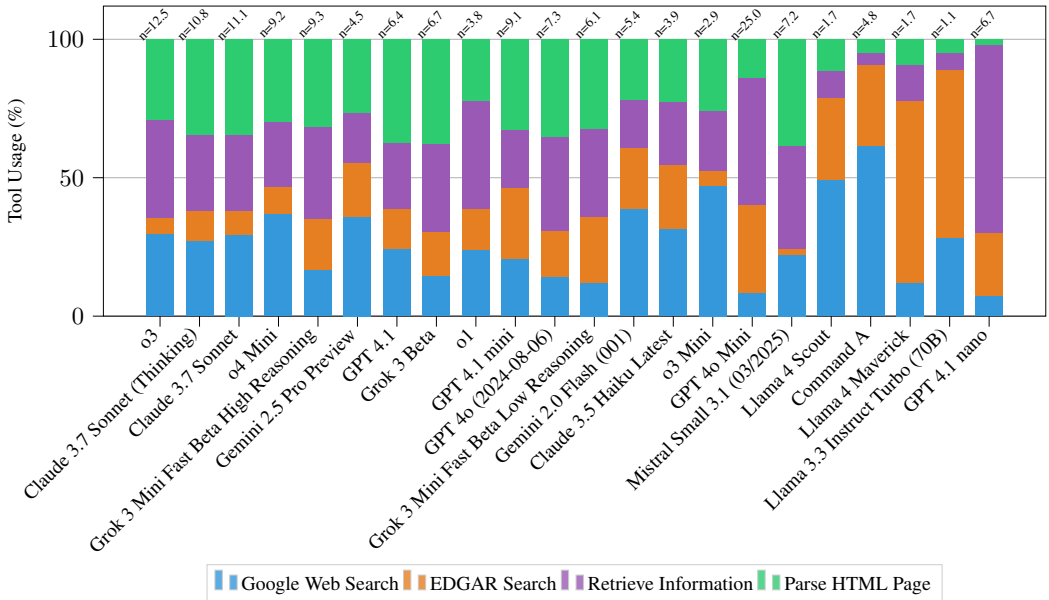}}
    \caption{Tool Usage Statistics}
    \label{fig:tool_usage_analysis_all}
\end{figure}

Figure \ref{fig:tool_usage_analysis_all} the average distribution of tool usage across the four available tools, for each model. In general, tool usage varied significantly by model. For example, mistral barely used the Edgar Search tool, whereas other models like LLaMA Maverick used it significantly.  Tool usage also varied significantly even between models from the same generation or provider - GPT 4.1 Nano uses significantly more RetrieveInformation calls than GPT 4.1. 

For the rest of the detailed metrics, we present the metrics described above categorized by question type. The results are displayed using two model subsets:

\begin{itemize}
    \item \textbf{Reasoning Models:} These are displayed separately due to their increasing prominence in recent research.
    \item \textbf{Full Range Subset:} This subset includes six models spanning the performance spectrum, from the highest-performing to one of the lowest-performing, with several intermediate performers included.
\end{itemize}

\subsubsection{Time and Cost Analysis}
\begin{figure}[h]  
    \centering
    \resizebox{\textwidth}{!}{
\begin{tikzpicture}

\definecolor{darkgray176}{RGB}{176,176,176}
\definecolor{darkorange25512714}{RGB}{255,127,14}
\definecolor{steelblue31119180}{RGB}{31,119,180}

\begin{axis}[
width=15cm,
height=8cm,
tick align=outside,
tick pos=left,
x grid style={darkgray176},
xmin=-1.54, xmax=23.54,
xtick style={color=black},
xtick={0,1,2,3,4,5,6,7,8,9,10,11,12,13,14,15,16,17,18,19,20,21,22},
xticklabel style={rotate=45.0,anchor=east},
xticklabels={
  Human Expert,
  o3,
  Claude 3.7 Sonnet (Thinking),
  Claude 3.7 Sonnet,
  o4 Mini,
  Grok 3 Mini Fast Beta High Reasoning,
  Gemini 2.5 Pro Preview,
  GPT 4.1,
  Grok 3 Beta,
  o1,
  GPT 4.1 mini,
  GPT 4o (2024-08-06),
  Grok 3 Mini Fast Beta Low Reasoning,
  Gemini 2.0 Flash (001),
  Claude 3.5 Haiku Latest,
  o3 Mini,
  GPT 4o Mini,
  Mistral Small 3.1 (03/2025),
  Llama 4 Scout,
  Command A,
  Llama 4 Maverick,
  Llama 3.3 Instruct Turbo (70B),
  GPT 4.1 nano
},
y grid style={darkgray176},
ylabel={Time per Question},
ymajorgrids,
ymin=0, ymax=21.8,
ytick style={color=black}
]
\draw[draw=none,fill=darkorange25512714,fill opacity=0.7] (axis cs:-0.4,0) rectangle (axis cs:0.4,16.8416352290824);
\draw[draw=none,fill=steelblue31119180,fill opacity=0.7] (axis cs:0.6,0) rectangle (axis cs:1.4,3.10898735223926);
\draw[draw=none,fill=steelblue31119180,fill opacity=0.7] (axis cs:1.6,0) rectangle (axis cs:2.4,2.52145450646082);
\draw[draw=none,fill=steelblue31119180,fill opacity=0.7] (axis cs:2.6,0) rectangle (axis cs:3.4,2.01611871324253);
\draw[draw=none,fill=steelblue31119180,fill opacity=0.7] (axis cs:3.6,0) rectangle (axis cs:4.4,2.74047992509801);
\draw[draw=none,fill=steelblue31119180,fill opacity=0.7] (axis cs:4.6,0) rectangle (axis cs:5.4,4.21628696511071);
\draw[draw=none,fill=steelblue31119180,fill opacity=0.7] (axis cs:5.6,0) rectangle (axis cs:6.4,1.20887063292423);
\draw[draw=none,fill=steelblue31119180,fill opacity=0.7] (axis cs:6.6,0) rectangle (axis cs:7.4,1.02183384658461);
\draw[draw=none,fill=steelblue31119180,fill opacity=0.7] (axis cs:7.6,0) rectangle (axis cs:8.4,1.00847122317596);
\draw[draw=none,fill=steelblue31119180,fill opacity=0.7] (axis cs:8.6,0) rectangle (axis cs:9.4,7.10817845990227);
\draw[draw=none,fill=steelblue31119180,fill opacity=0.7] (axis cs:9.6,0) rectangle (axis cs:10.4,0.860051496497978);
\draw[draw=none,fill=steelblue31119180,fill opacity=0.7] (axis cs:10.6,0) rectangle (axis cs:11.4,0.680976663021744);
\draw[draw=none,fill=steelblue31119180,fill opacity=0.7] (axis cs:11.6,0) rectangle (axis cs:12.4,1.34165432394325);
\draw[draw=none,fill=steelblue31119180,fill opacity=0.7] (axis cs:12.6,0) rectangle (axis cs:13.4,0.392756475497993);
\draw[draw=none,fill=steelblue31119180,fill opacity=0.7] (axis cs:13.6,0) rectangle (axis cs:14.4,0.78061433336557);
\draw[draw=none,fill=steelblue31119180,fill opacity=0.7] (axis cs:14.6,0) rectangle (axis cs:15.4,2.43564873369331);
\draw[draw=none,fill=steelblue31119180,fill opacity=0.7] (axis cs:15.6,0) rectangle (axis cs:16.4,1.55678914967652);
\draw[draw=none,fill=steelblue31119180,fill opacity=0.7] (axis cs:16.6,0) rectangle (axis cs:17.4,0.651606206961423);
\draw[draw=none,fill=steelblue31119180,fill opacity=0.7] (axis cs:17.6,0) rectangle (axis cs:18.4,0.23031322903792);
\draw[draw=none,fill=steelblue31119180,fill opacity=0.7] (axis cs:18.6,0) rectangle (axis cs:19.4,1.58207727882719);
\draw[draw=none,fill=steelblue31119180,fill opacity=0.7] (axis cs:19.6,0) rectangle (axis cs:20.4,0.196181884822885);
\draw[draw=none,fill=steelblue31119180,fill opacity=0.7] (axis cs:20.6,0) rectangle (axis cs:21.4,0.0557774186148594);
\draw[draw=none,fill=steelblue31119180,fill opacity=0.7] (axis cs:21.6,0) rectangle (axis cs:22.4,1.09160743126867);
\draw (axis cs:0,16.8416352290824) ++(0pt,0pt) node[
  scale=0.55,
  anchor=south west,
  text=black,
  rotate=70
]{1010.5s (16.8m)};
\draw (axis cs:1,3.10898735223926) ++(0pt,0pt) node[
  scale=0.55,
  anchor=south west,
  text=black,
  rotate=70
]{186.5s (3.1m)};
\draw (axis cs:2,2.52145450646082) ++(0pt,0pt) node[
  scale=0.55,
  anchor=south west,
  text=black,
  rotate=70
]{151.3s (2.5m)};
\draw (axis cs:3,2.01611871324253) ++(0pt,0pt) node[
  scale=0.55,
  anchor=south west,
  text=black,
  rotate=70
]{121.0s (2.0m)};
\draw (axis cs:4,2.74047992509801) ++(0pt,0pt) node[
  scale=0.55,
  anchor=south west,
  text=black,
  rotate=70
]{164.4s (2.7m)};
\draw (axis cs:5,4.21628696511071) ++(0pt,0pt) node[
  scale=0.55,
  anchor=south west,
  text=black,
  rotate=70
]{253.0s (4.2m)};
\draw (axis cs:6,1.20887063292423) ++(0pt,0pt) node[
  scale=0.55,
  anchor=south west,
  text=black,
  rotate=70
]{72.5s (1.2m)};
\draw (axis cs:7,1.02183384658461) ++(0pt,0pt) node[
  scale=0.55,
  anchor=south west,
  text=black,
  rotate=70
]{61.3s (1.0m)};
\draw (axis cs:8,1.00847122317596) ++(0pt,0pt) node[
  scale=0.55,
  anchor=south west,
  text=black,
  rotate=70
]{60.5s (1.0m)};
\draw (axis cs:9,7.10817845990227) ++(0pt,0pt) node[
  scale=0.55,
  anchor=south west,
  text=black,
  rotate=70
]{426.5s (7.1m)};
\draw (axis cs:10,0.860051496497978) ++(0pt,0pt) node[
  scale=0.55,
  anchor=south west,
  text=black,
  rotate=70
]{51.6s (0.9m)};
\draw (axis cs:11,0.680976663021744) ++(0pt,0pt) node[
  scale=0.55,
  anchor=south west,
  text=black,
  rotate=70
]{40.9s (0.7m)};
\draw (axis cs:12,1.34165432394325) ++(0pt,0pt) node[
  scale=0.55,
  anchor=south west,
  text=black,
  rotate=70
]{80.5s (1.3m)};
\draw (axis cs:13,0.392756475497993) ++(0pt,0pt) node[
  scale=0.55,
  anchor=south west,
  text=black,
  rotate=70
]{23.6s (0.4m)};
\draw (axis cs:14,0.78061433336557) ++(0pt,0pt) node[
  scale=0.55,
  anchor=south west,
  text=black,
  rotate=70
]{46.8s (0.8m)};
\draw (axis cs:15,2.43564873369331) ++(0pt,0pt) node[
  scale=0.55,
  anchor=south west,
  text=black,
  rotate=70
]{146.1s (2.4m)};
\draw (axis cs:16,1.55678914967652) ++(0pt,0pt) node[
  scale=0.55,
  anchor=south west,
  text=black,
  rotate=70
]{93.4s (1.6m)};
\draw (axis cs:17,0.651606206961423) ++(0pt,0pt) node[
  scale=0.55,
  anchor=south west,
  text=black,
  rotate=70
]{39.1s (0.7m)};
\draw (axis cs:18,0.23031322903792) ++(0pt,0pt) node[
  scale=0.55,
  anchor=south west,
  text=black,
  rotate=70
]{13.8s (0.2m)};
\draw (axis cs:19,1.58207727882719) ++(0pt,0pt) node[
  scale=0.55,
  anchor=south west,
  text=black,
  rotate=70
]{94.9s (1.6m)};
\draw (axis cs:20,0.196181884822885) ++(0pt,0pt) node[
  scale=0.55,
  anchor=south west,
  text=black,
  rotate=70
]{11.8s (0.2m)};
\draw (axis cs:21,0.0557774186148594) ++(0pt,0pt) node[
  scale=0.55,
  anchor=south west,
  text=black,
  rotate=70
]{3.3s (0.1m)};
\draw (axis cs:22,1.09160743126867) ++(0pt,0pt) node[
  scale=0.55,
  anchor=south west,
  text=black,
  rotate=70
]{65.5s (1.1m)};
\end{axis}

\end{tikzpicture}}
    \caption{Overall Time Analysis Compared to Expert}
    \label{fig:time_analysis_all}
\end{figure}
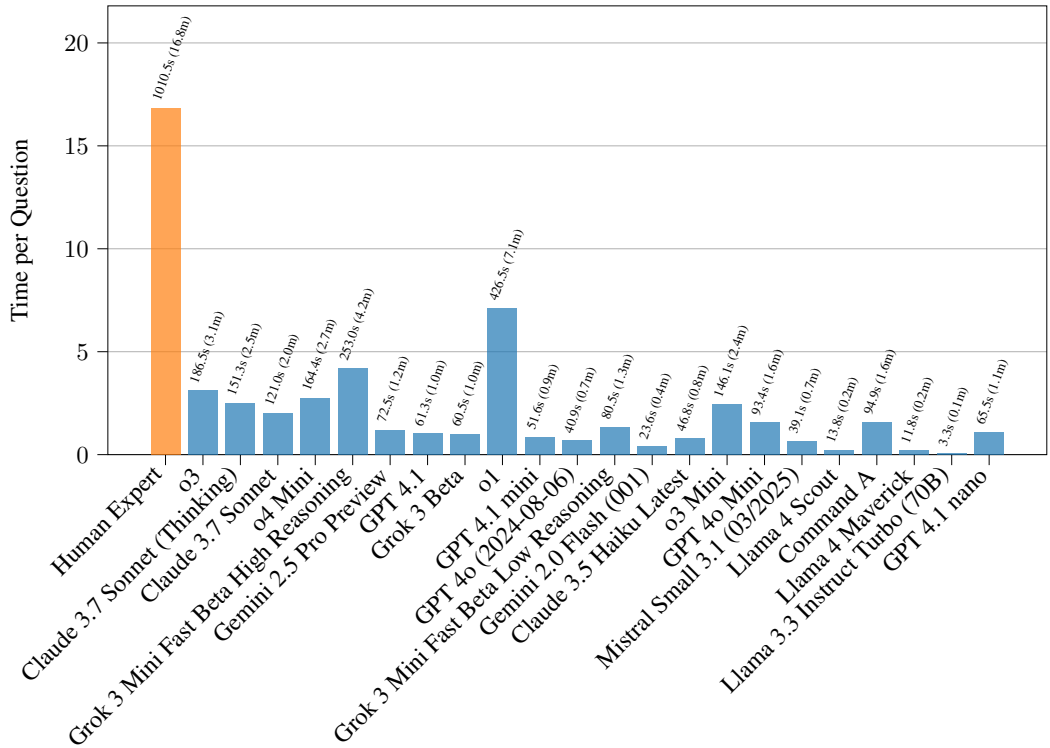

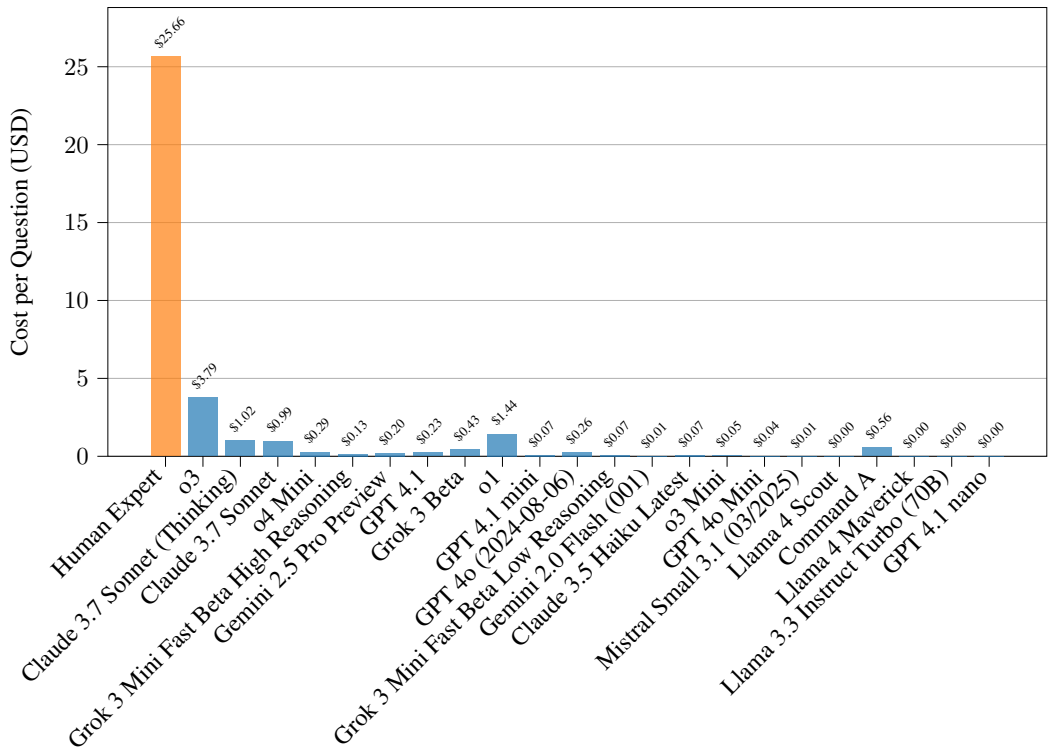
\begin{figure}[h]  
    \begin{centering}
    \resizebox{\textwidth}{!}{
\begin{tikzpicture}

\definecolor{darkgray176}{RGB}{176,176,176}
\definecolor{darkorange25512714}{RGB}{255,127,14}
\definecolor{steelblue31119180}{RGB}{31,119,180}

\begin{axis}[
width=15cm,
height=8cm,
tick align=outside,
tick pos=left,
x grid style={darkgray176},
xmin=-1.54, xmax=23.54,
xtick style={color=black},
xtick={0,1,2,3,4,5,6,7,8,9,10,11,12,13,14,15,16,17,18,19,20,21,22},
xticklabel style={rotate=45.0,anchor=east},
xticklabels={
  Human Expert,
  o3,
  Claude 3.7 Sonnet (Thinking),
  Claude 3.7 Sonnet,
  o4 Mini,
  Grok 3 Mini Fast Beta High Reasoning,
  Gemini 2.5 Pro Preview,
  GPT 4.1,
  Grok 3 Beta,
  o1,
  GPT 4.1 mini,
  GPT 4o (2024-08-06),
  Grok 3 Mini Fast Beta Low Reasoning,
  Gemini 2.0 Flash (001),
  Claude 3.5 Haiku Latest,
  o3 Mini,
  GPT 4o Mini,
  Mistral Small 3.1 (03/2025),
  Llama 4 Scout,
  Command A,
  Llama 4 Maverick,
  Llama 3.3 Instruct Turbo (70B),
  GPT 4.1 nano
},
y grid style={darkgray176},
ylabel={Cost per Question (USD)},
ymajorgrids,
ymin=0, ymax=28.8,
ytick style={color=black}
]
\draw[draw=none,fill=darkorange25512714,fill opacity=0.7] (axis cs:-0.4,0) rectangle (axis cs:0.4,25.6649679255987);
\draw[draw=none,fill=steelblue31119180,fill opacity=0.7] (axis cs:0.6,0) rectangle (axis cs:1.4,3.78610164342291);
\draw[draw=none,fill=steelblue31119180,fill opacity=0.7] (axis cs:1.6,0) rectangle (axis cs:2.4,1.01676200923806);
\draw[draw=none,fill=steelblue31119180,fill opacity=0.7] (axis cs:2.6,0) rectangle (axis cs:3.4,0.988625833099283);
\draw[draw=none,fill=steelblue31119180,fill opacity=0.7] (axis cs:3.6,0) rectangle (axis cs:4.4,0.2862863155701);
\draw[draw=none,fill=steelblue31119180,fill opacity=0.7] (axis cs:4.6,0) rectangle (axis cs:5.4,0.13049179427344);
\draw[draw=none,fill=steelblue31119180,fill opacity=0.7] (axis cs:5.6,0) rectangle (axis cs:6.4,0.196306608510841);
\draw[draw=none,fill=steelblue31119180,fill opacity=0.7] (axis cs:6.6,0) rectangle (axis cs:7.4,0.230873824907237);
\draw[draw=none,fill=steelblue31119180,fill opacity=0.7] (axis cs:7.6,0) rectangle (axis cs:8.4,0.425365855975696);
\draw[draw=none,fill=steelblue31119180,fill opacity=0.7] (axis cs:8.6,0) rectangle (axis cs:9.4,1.43975413124239);
\draw[draw=none,fill=steelblue31119180,fill opacity=0.7] (axis cs:9.6,0) rectangle (axis cs:10.4,0.0683812668820963);
\draw[draw=none,fill=steelblue31119180,fill opacity=0.7] (axis cs:10.6,0) rectangle (axis cs:11.4,0.257491518299961);
\draw[draw=none,fill=steelblue31119180,fill opacity=0.7] (axis cs:11.6,0) rectangle (axis cs:12.4,0.0666693135255327);
\draw[draw=none,fill=steelblue31119180,fill opacity=0.7] (axis cs:12.6,0) rectangle (axis cs:13.4,0.0114746008157894);
\draw[draw=none,fill=steelblue31119180,fill opacity=0.7] (axis cs:13.6,0) rectangle (axis cs:14.4,0.0664669898632701);
\draw[draw=none,fill=steelblue31119180,fill opacity=0.7] (axis cs:14.6,0) rectangle (axis cs:15.4,0.0471560993470098);
\draw[draw=none,fill=steelblue31119180,fill opacity=0.7] (axis cs:15.6,0) rectangle (axis cs:16.4,0.0374871363283261);
\draw[draw=none,fill=steelblue31119180,fill opacity=0.7] (axis cs:16.6,0) rectangle (axis cs:17.4,0.00605787514245645);
\draw[draw=none,fill=steelblue31119180,fill opacity=0.7] (axis cs:17.6,0) rectangle (axis cs:18.4,0.00459797897284824);
\draw[draw=none,fill=steelblue31119180,fill opacity=0.7] (axis cs:18.6,0) rectangle (axis cs:19.4,0.562849864674662);
\draw[draw=none,fill=steelblue31119180,fill opacity=0.7] (axis cs:19.6,0) rectangle (axis cs:20.4,0.00233937247067305);
\draw[draw=none,fill=steelblue31119180,fill opacity=0.7] (axis cs:20.6,0) rectangle (axis cs:21.4,0.00302138270884077);
\draw[draw=none,fill=steelblue31119180,fill opacity=0.7] (axis cs:21.6,0) rectangle (axis cs:22.4,0.00318040528592362);
\draw (axis cs:0,25.6649679255987) ++(-2pt,0pt) node[
  scale=0.55,
  anchor=south west,
  text=black,
  rotate=45
]{\$25.66};
\draw (axis cs:1,3.78610164342291) ++(-2pt,0pt) node[
  scale=0.55,
  anchor=south west,
  text=black,
  rotate=45
]{\$3.79};
\draw (axis cs:2,1.01676200923806) ++(-2pt,0pt) node[
  scale=0.55,
  anchor=south west,
  text=black,
  rotate=45
]{\$1.02};
\draw (axis cs:3,0.988625833099283) ++(-2pt,0pt) node[
  scale=0.55,
  anchor=south west,
  text=black,
  rotate=45
]{\$0.99};
\draw (axis cs:4,0.2862863155701) ++(-2pt,0pt) node[
  scale=0.55,
  anchor=south west,
  text=black,
  rotate=45
]{\$0.29};
\draw (axis cs:5,0.13049179427344) ++(-2pt,0pt) node[
  scale=0.55,
  anchor=south west,
  text=black,
  rotate=45
]{\$0.13};
\draw (axis cs:6,0.196306608510841) ++(-2pt,0pt) node[
  scale=0.55,
  anchor=south west,
  text=black,
  rotate=45
]{\$0.20};
\draw (axis cs:7,0.230873824907237) ++(-2pt,0pt) node[
  scale=0.55,
  anchor=south west,
  text=black,
  rotate=45
]{\$0.23};
\draw (axis cs:8,0.425365855975696) ++(-2pt,0pt) node[
  scale=0.55,
  anchor=south west,
  text=black,
  rotate=45
]{\$0.43};
\draw (axis cs:9,1.43975413124239) ++(-2pt,0pt) node[
  scale=0.55,
  anchor=south west,
  text=black,
  rotate=45
]{\$1.44};
\draw (axis cs:10,0.0683812668820963) ++(-2pt,0pt) node[
  scale=0.55,
  anchor=south west,
  text=black,
  rotate=45
]{\$0.07};
\draw (axis cs:11,0.257491518299961) ++(-2pt,0pt) node[
  scale=0.55,
  anchor=south west,
  text=black,
  rotate=45
]{\$0.26};
\draw (axis cs:12,0.0666693135255327) ++(-2pt,0pt) node[
  scale=0.55,
  anchor=south west,
  text=black,
  rotate=45
]{\$0.07};
\draw (axis cs:13,0.0114746008157894) ++(-2pt,0pt) node[
  scale=0.55,
  anchor=south west,
  text=black,
  rotate=45
]{\$0.01};
\draw (axis cs:14,0.0664669898632701) ++(-2pt,0pt) node[
  scale=0.55,
  anchor=south west,
  text=black,
  rotate=45
]{\$0.07};
\draw (axis cs:15,0.0471560993470098) ++(-2pt,0pt) node[
  scale=0.55,
  anchor=south west,
  text=black,
  rotate=45
]{\$0.05};
\draw (axis cs:16,0.0374871363283261) ++(-2pt,0pt) node[
  scale=0.55,
  anchor=south west,
  text=black,
  rotate=45
]{\$0.04};
\draw (axis cs:17,0.00605787514245645) ++(-2pt,0pt) node[
  scale=0.55,
  anchor=south west,
  text=black,
  rotate=45
]{\$0.01};
\draw (axis cs:18,0.00459797897284824) ++(-2pt,0pt) node[
  scale=0.55,
  anchor=south west,
  text=black,
  rotate=45
]{\$0.00};
\draw (axis cs:19,0.562849864674662) ++(-2pt,0pt) node[
  scale=0.55,
  anchor=south west,
  text=black,
  rotate=45
]{\$0.56};
\draw (axis cs:20,0.00233937247067305) ++(-2pt,0pt) node[
  scale=0.55,
  anchor=south west,
  text=black,
  rotate=45
]{\$0.00};
\draw (axis cs:21,0.00302138270884077) ++(-2pt,0pt) node[
  scale=0.55,
  anchor=south west,
  text=black,
  rotate=45
]{\$0.00};
\draw (axis cs:22,0.00318040528592362) ++(-2pt,0pt) node[
  scale=0.55,
  anchor=south west,
  text=black,
  rotate=45
]{\$0.00};
\end{axis}

\end{tikzpicture}}
    \caption{Overall Cost Analysis Compared to Expert (avg \$91.4 /h)}
    \label{fig:cost_analysis_all}
    \end{centering}
\end{figure}

The overall takeaway from Figure \ref{fig:time_analysis_all} and Figure \ref{fig:cost_analysis_all} is that all models are significantly more time effective than human experts, ranging from two times faster to more than ten times faster.

\subsection{Tool Call Analysis by Question Type: Figure \ref{fig:reaosning_models_tool_calls_q_type}, Figure \ref{fig:full_range_models_tool_calls_q_type} and Figure \ref{fig:tool_calling_all}}

\begin{figure}[h]  
    \centering
    \resizebox{\textwidth}{!}{\input{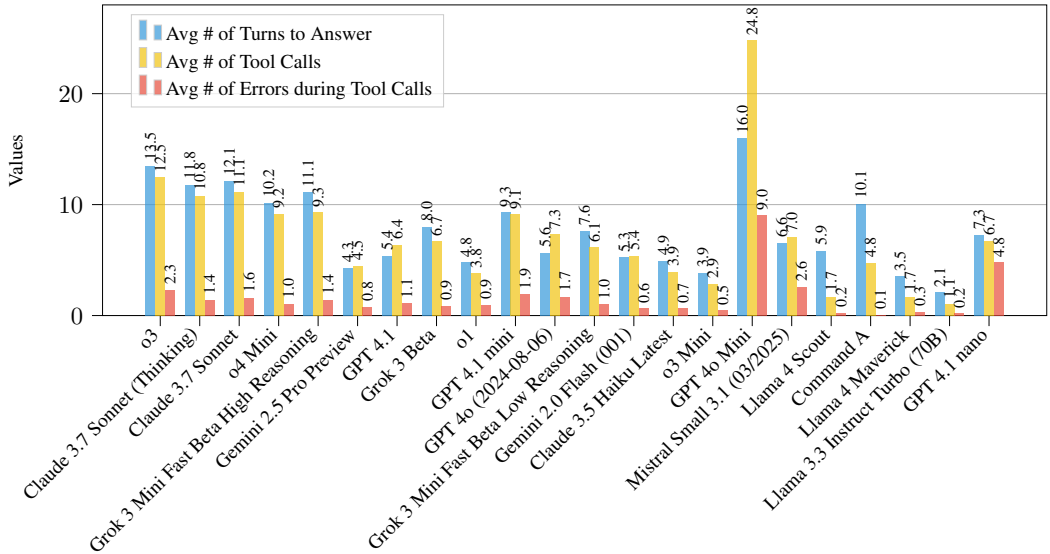}}
    \caption{Tool Usage Statistics on All Models}
    \label{fig:tool_calling_all}
\end{figure}

\begin{figure}[h]  
    \centering
    \resizebox{\textwidth}{!}{\input{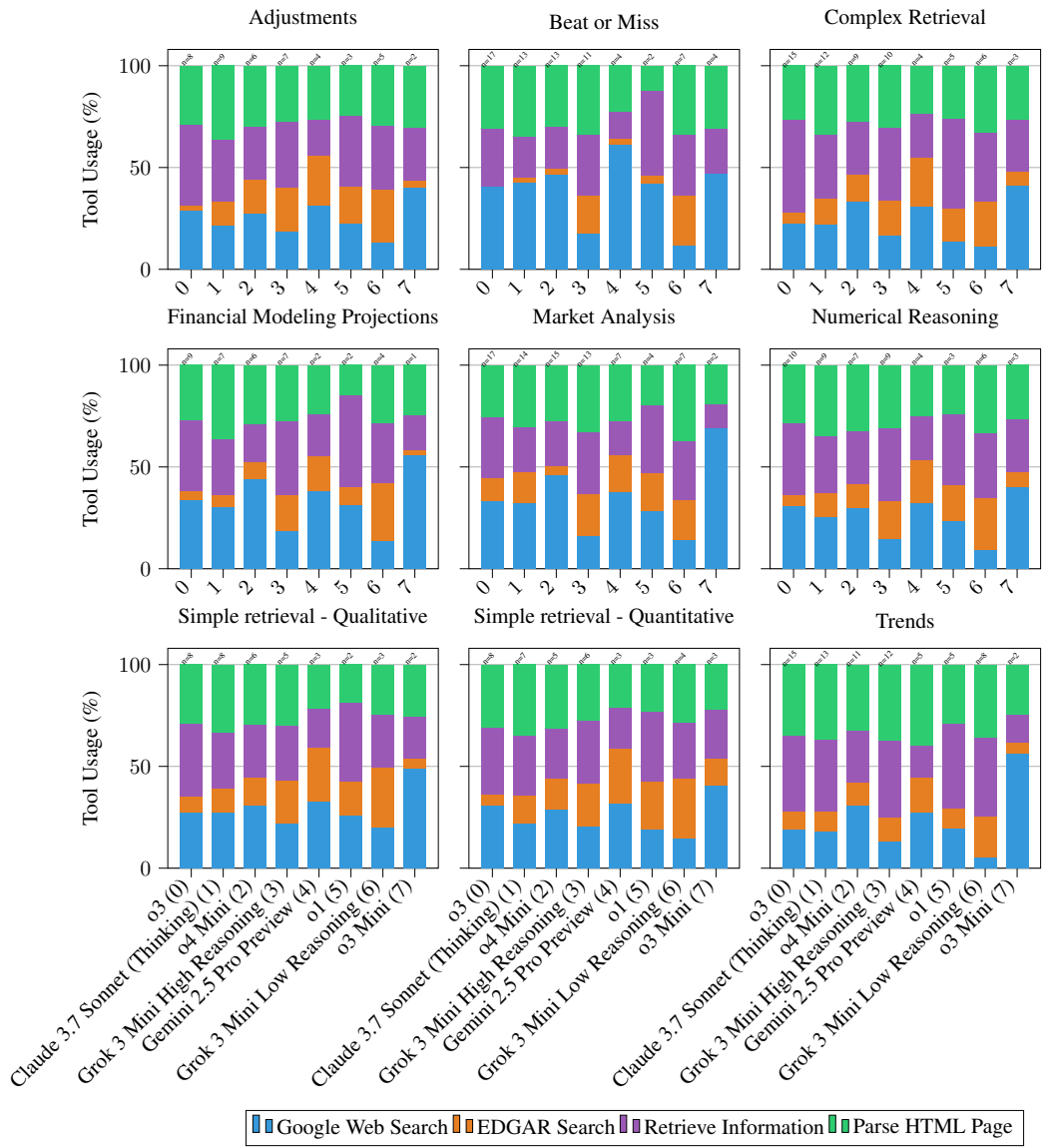}}
    \caption{Reasoning Models Tool Usage by Question Type}
    \label{fig:reaosning_models_tool_calls_q_type}
\end{figure}

\begin{figure}[h]  
    \centering
    \resizebox{\textwidth}{!}{\input{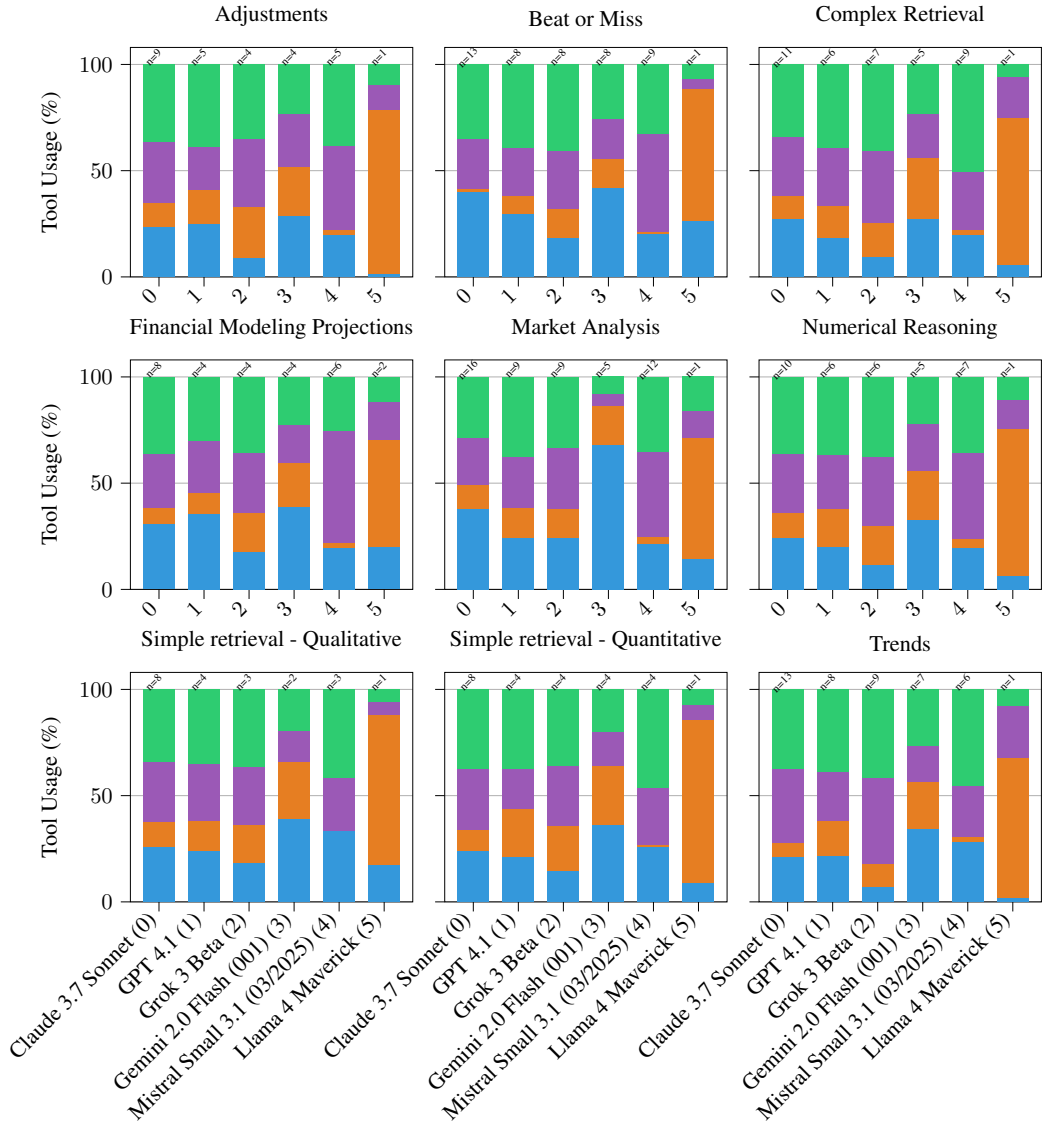}}
    \caption{Full Range Subset Models Tool Usage by Question Type}
    \label{fig:full_range_models_tool_calls_q_type}
\end{figure}

\subsection{Time Analysis by Question Type: Figure \ref{fig:reaosning_models_time_q_type} and Figure \ref{fig:full_range_models_time_q_type}}

\begin{figure}[h]  
    \centering
    \resizebox{\textwidth}{!}{\input{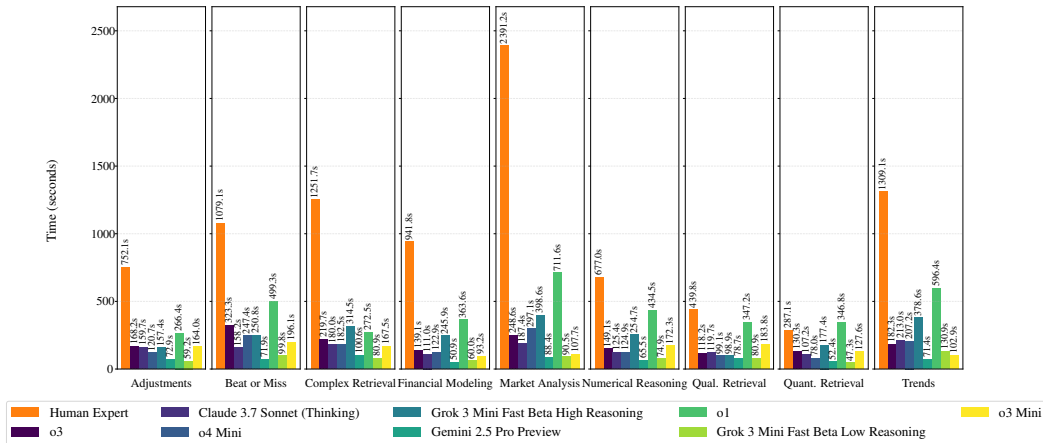}}
    \caption{Reasoning Models Time by Question Type}
    \label{fig:reaosning_models_time_q_type}
\end{figure}

\begin{figure}[h]  
    \centering
    \resizebox{\textwidth}{!}{\input{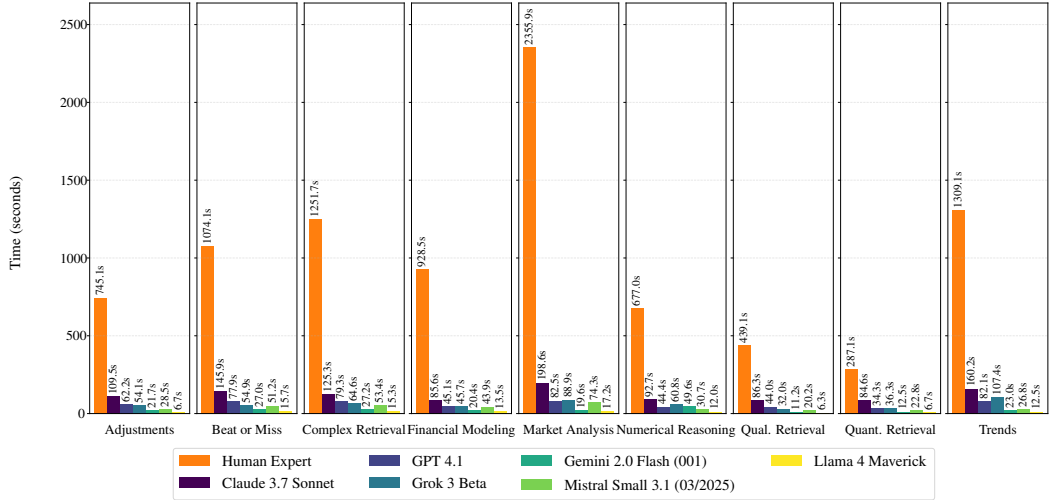}}
    \caption{Full Range Subset Models Time by Question Type}
    \label{fig:full_range_models_time_q_type}
\end{figure}

\subsection{Cost Analysis by Question Type: Figure \ref{fig:reaosning_models_cost_q_type} and Figure \ref{fig:full_range_models_cost_q_type}}

\begin{figure}[h]  
    \centering
    \resizebox{\textwidth}{!}{\input{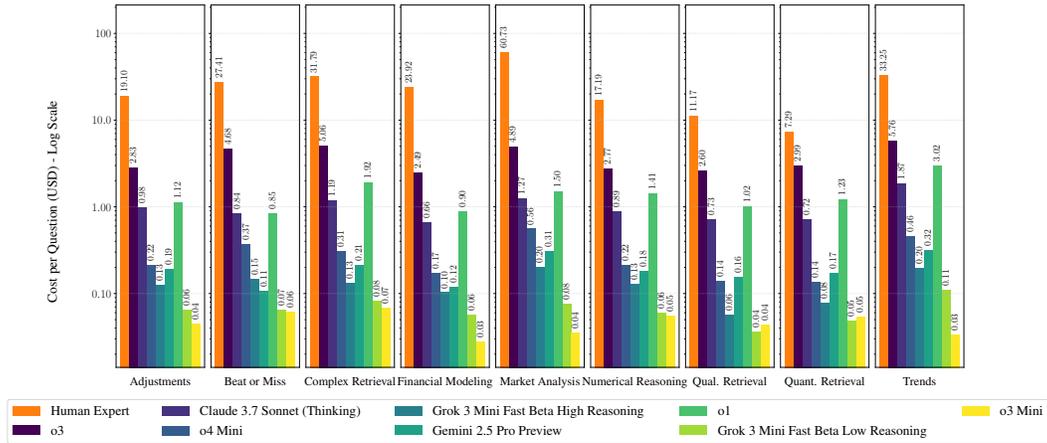}}
    \caption{Reasoning Models Cost by Question Type}
    \label{fig:full_range_models_cost_q_type}
\end{figure}

\begin{figure}[h]  
    \centering
    \resizebox{\textwidth}{!}{\input{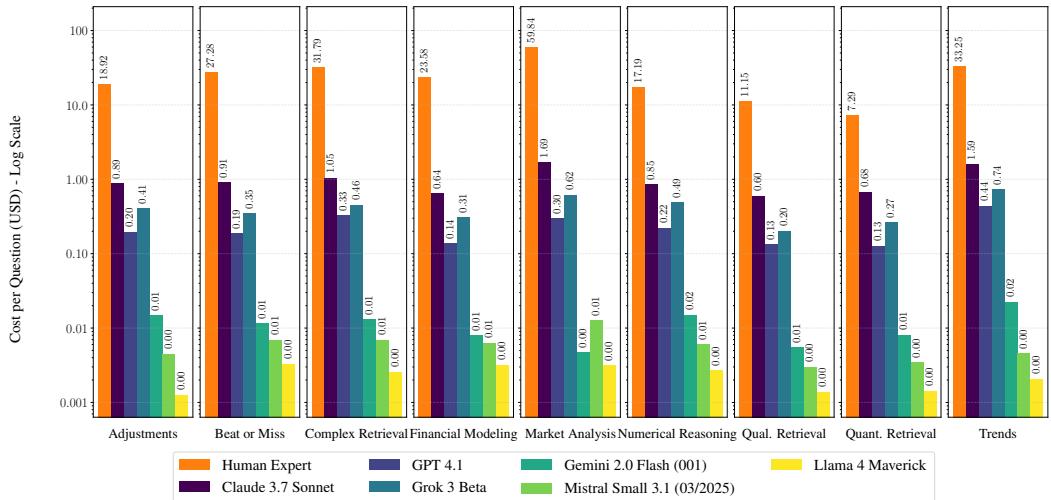}}
    \caption{Full Range Subset Models Cost by Question Type}
    \label{fig:full_range_models_cost_q_type}
\end{figure}

\end{appendices}

\end{document}